\documentclass[useAMS,usenatbib]{mn2e}

\include{epsf}
\usepackage{rotating}

\def\gsim{\lower.73ex\hbox{$\sim$}\llap{\raise.4ex\hbox{$>$}}$\,$}
\def\lsim{\lower.73ex\hbox{$\sim$}\llap{\raise.4ex\hbox{$<$}}$\,$}

\def\kms{km s$^{-1}$}
\def\Msun{M_{\odot}}
\def\Mpch{\, h^{-1}{\rm Mpc}}

\def\kms{\, {{\rm km}}\,{{\rm s}}^{-1} }
\def\be{\begin{equation}}
\def\ee{\end{equation}}

\title[The 2SLAQ survey: QSO clustering and the L-$z$ degeneracy]
  {The 2dF-SDSS LRG and QSO survey: QSO clustering and the L-$z$ degeneracy}
\author[J. da \^{A}ngela et~al.]
  {J.~da \^{A}ngela$^{1}$, T.~Shanks$^1$, S.~M.~Croom$^{2}$, P.~Weilbacher$^{3}$,  R.~J.~Brunner$^{4,5}$,
\newauthor W.J. Couch$^6$, L.~Miller$^{7}$, A.~D.~Myers$^{4,5}$, R.~C.~Nichol$^8$, K.~A.~Pimbblet$^{9}$,
\newauthor R.~de Propris$^{10}$, G.~T.~Richards$^{11}$, N.~P.~Ross$^{1}$, D.~P.~Schneider$^{12}$, D.~A.~Wake$^{1}$  \\
1 Department of Physics, Durham University, Science Laboratories, South Road, Durham, DH1 3LE, United Kingdom\\
2 Anglo-Australian Telescope, PO Box 296, NSW 1710, Australia\\
3 Astrophysikalisches Institut Potsdam, An der Sternwarte 16, D-14482 Potsdam, Germany\\
4 Department of Astronomy, University of Illinois at Urbana-Champaign, Urbana, IL 61801 USA\\
5 National Center for Supercomputing Applications, Champaign, IL 61820 USA\\
6 Centre for Astrophysics \& Supercomputing, Applied Science Building, Swinburne University, Hawthorn VIC 3122, Australia \\ 
7 Dept. of Physics, University of Oxford, Denys Wilkinson Building, Keble Road, Oxford OX1 3RH, UK\\
8 Institute of Cosmology and Gravitation, Mercantile House, University of Portsmouth,  Portsmouth, PO1 2EG, UK \\
9 Department of Physics, University of Queensland, Brisbane, QLD 4072, Australia\\
10 Cerro Tololo Inter-American Observatory, Casilla 603, La Serena, Chile\\
11 Department of Physics, Drexel University,Philadelphia, PA, 19104 USA\\
12 Department of Astronomy, the Pennsylvania State University, University Park, PA 16802 USA\\
}

\begin{document}

\pagerange{\pageref{firstpage}--\pageref{lastpage}} \pubyear{2006}

\maketitle

\label{firstpage}

\begin{abstract}

We combine the QSO samples from the 2dF QSO Redshift Survey
(2QZ) and the 2dF-SDSS LRG and QSO Survey (2SLAQ) in order to
investigate the clustering of $z\sim 1.4$ QSOs and measure the
correlation function ($\xi$). The clustering signal in redshift-space
and projected along the sky direction is similar to that previously
obtained from the 2QZ sample alone. By fitting functional forms to
$\xi(\sigma,\pi)$, the correlation function  measured along and across
the line of sight, we find, as expected,  that $\beta$,  the dynamical
infall parameter and $\Omega_{m}^{0}$, the cosmological density
parameter, are degenerate. However, this degeneracy can be lifted by
using linear theory predictions under different cosmological scenarios.
Using the combination of the 2QZ and 2SLAQ QSO data, we obtain:
$\beta_{QSO}(z=1.4) = 0.60^{+0.14}_{-0.11}$,
$\Omega_{m}^{0}=0.25_{-0.07}^{+0.09}$ which imply a value for the QSO
bias, $b(z=1.4)=1.5\pm0.2$.

The combination of the 2QZ with the fainter 2SLAQ QSO sample further
reveals that QSO clustering does not depend strongly on luminosity at
fixed redshift. This result is inconsistent with the expectation of
simple `high peaks' biasing models where more luminous, rare QSOs are
assumed to inhabit higher mass haloes. The data are more  consistent
with models which predict that QSOs of different luminosities reside in
haloes of similar mass. By assuming ellipsoidal models for the collapse
of density perturbations, we estimate the mass of the dark matter haloes
which the QSOs inhabit. We find that halo mass does not evolve strongly
with redshift nor depend on QSO luminosity. Assuming a range of
relations which relate halo to black hole mass we investigate
how black hole mass correlates with luminosity and redshift and ascertain
the relation between Eddington efficiency and black hole mass. Our
results suggest that QSOs of different luminosities may contain black
holes of similar mass.

\end{abstract}

\begin{keywords}
 
surveys -  quasars, quasars: general, large-scale structure of Universe, cosmology: observations

\end{keywords}

\section{Introduction}

There is a significant amount of observational evidence for the
existence of supermassive black holes in the centre of galactic haloes.
This conclusion is based on studies which span a wide redshift-range.
Whilst at low-$z$, the evidence for the presence of black holes comes
from dynamical surveys of galaxies in the local Universe
\citep{kormendy05,richstone98,magorrian}, at high-$z$, black hole --
host galaxy studies are pursued by using the width of quasar (QSO) broad
emission lines to estimate black hole masses and the host galaxy's
narrow emission lines to determine stellar velocity dispersion
\citep[e.g.][]{shields06b}. These results hint at a correlation between
the growth/physics of the bulge and dark matter halo, and the physics of
accretion of mass onto the central black hole and subsequent growth
\citep[e.g.][]{tremaine02}. The relation between the bulge and its black
hole is the subject of intense observational and theoretical interest
\citep{kh,ferrarese_merrit,gebhardt,ferrarese,wyithe05}. Many
uncertainties still exist when trying to interpret this black hole -
bulge connection. One possible scenario is that the mechanism that
``feeds'' black hole growth is the same, or is correlated to, those
properties responsible for bulge growth, such as mergers or
instabilities, which may also lead to enhanced star formation; some of
the gas may instead ``fuel'' the  black hole, and consequently lead to
QSO activity \citep[e.g.][]{bower05}. This picture is supported by the
similar ``shape'' of the cosmological star formation history of the
Universe and the evolution of the QSO number density as a function of
redshift \citep[e.g.][]{schmidt70,bsp88,schmidt95,madau,dunlop}.

In the standard scenario, QSO activity is triggered by accretion onto a
supermassive black hole \citep[SMBH, e.g.][]{hopkins}. Given that the
growth of the SMBH relates to that of the underlying dark matter halo
\citep{baes03,wyithe,wyithe06} and the halo properties are correlated
with the local density contrast, clustering measurements provide an
insight into QSO and black hole physics.

QSO clustering measurements allow determinations of halo masses and how
they relate to black hole mass. QSO lifetimes, which have been the basis
of interpretations of QSO luminosity functions \citep{hopkins05} can
also be inferred from clustering measurements \citep[e.g.][]{scottnew},
and hence permit discrimination between QSO evolutionary models, such as
a cosmologically long-lived population \citep[e.g.][]{boyle}.
\cite{lance2} addressed the change of accretion efficiency with
redshift, arguing that, even though the mass of the black holes grows
with time as galaxies grow hierarchically, the mean accretion rate
decreases with decreasing redshift, hence leading to a decrease of the
QSO luminosity with time. This picture is supported by theoretical
models, such as that of \cite{kh}.

The evolution of QSO clustering has been the subject of recent studies.
In particular, the wealth of information contained in the Sloan Digital
Sky Survey (SDSS; \citealt{york_sdss}) and the 2dF QSO Redshift Survey
(2QZ; \citealt{scott04}) data have allowed studies such as those of
\cite{porciani}, \cite{myers06}, and \cite{scottnew}, who measured the
redshift dependence of QSO clustering. In particular, the latter
inferred the evolution of halo mass with redshift, besides estimating
black hole masses and accretion efficiencies, based on QSO clustering
measurements from the 2QZ sample. However, and as pointed out by those
authors, these studies do not take into account any potential luminosity
dependence of QSO clustering.

It is not trivial to address the possible dependence of QSO clustering
on luminosity. Due to the evolution of the QSO luminosity function  and
the flux-limited nature of the 2QZ and most other surveys, the most
luminous QSOs lie at high redshifts, while the faintest ones have low
redshifts. The lowest and highest redshift objects in the 2QZ sample
extend throughout separate luminosity ranges, hence hampering any
attempt to study the effects of luminosity on QSO clustering, black hole
masses and accretion efficiencies, free from any possible evolutionary
biases.

This necessary caveat in any study of luminosity dependence of QSO
clustering was one of the main motivations for the 2SLAQ (2dF-SDSS LRG
and QSO) QSO survey. Using faint, photometric QSO candidates from the
SDSS QSO survey, the observations at the 2dF facility result in an
extension of the previous 2QZ survey to fainter magnitudes. The faint
magnitude limit of $g = 21.85$ is $\sim 1$ magnitude fainter than that
of the 2QZ, and the new data, spanning a similar $z$-range as the 2QZ,
constitute a new, potentially powerful tool to disentangle the effects
of luminosity and redshift on the clustering of QSOs, thus providing a
new test of current QSO, black hole and bias models. With its fainter
QSO magnitude limit than the photometric SDSS QSO catalogue of
\cite{myers06}, 2SLAQ, despite its smaller statistical weight, 
constitutes a more valuable tool for breaking the $L$-$z$ degeneracy.

In this paper we combine the 2QZ and 2SLAQ QSO samples and analyse the
clustering of $z\sim 1.5$ QSOs. In addition, we use the wide luminosity
range covered by the combination of the two ensembles to determine the
luminosity dependence of QSO clustering, free from evolutionary effects.
In section \ref{sec:2slaq} we present a brief description of the 2SLAQ
QSO survey. We then measure the clustering signal of the QSOs, in
redshift-space  ($z$-space); projected along the sky direction (and
hence free of dynamical distortions); and in orthogonal directions
(section \ref{sec:clust}). These measurements allow us to model the
anisotropies due to dynamical and geometrical distortions in the
clustering signal and constrain $\Omega_{m}^{0}$ and
$\beta_{QSO}(z=1.5)$. This analysis is discussed in section
\ref{sec:zdist}. In  section \ref{sec:Lz} we describe the L-$z$
degeneracy and how we attempt to break it by combining the 2QZ and 2SLAQ
QSO samples. Our QSO clustering measurements as a function of magnitude
and redshift follow in section \ref{sec:Lz_xi}. We then attempt to
determine if QSO bias correlates with QSO luminosity, and how these
results affect the average mass of the dark matter haloes the QSO
inhabit (section \ref{sec:Lz_bias}). Assuming that the mass of the dark
matter halo correlates with that of the black hole associated to the
QSO, we determine how the black hole mass changes with redshift and
luminosity, and discuss how our results affect the black hole accretion
efficiency, in section \ref{sec:Lz_massbh}. Finally, in section
\ref{sec:conc}, we outline the conclusions of this paper.

\section{The 2SLAQ QSO Survey}
\label{sec:2slaq}

The 2SLAQ QSO survey is an extension of the previous 2QZ survey to
fainter magnitudes. The main aspects and description of this survey can
be found in \cite{richards04}, who report on the first 3 semesters of
the data collection and present luminosity function results from the
sample of $\sim 5600$ QSOs obtained at the time. Now that the survey has
been completed and the analysis of the data is being developed, there
are a total of $\sim 9000$ ($z\lsim 3$) QSOs. Both the imaging and
spectroscopic data, obtained from  the Sloan telescope and  the AAT
respectively, are extensively described by Croom et al., (in prep.).

\begin{figure*}
\begin{center}
\centerline{\epsfxsize = 17.0cm
\epsfbox{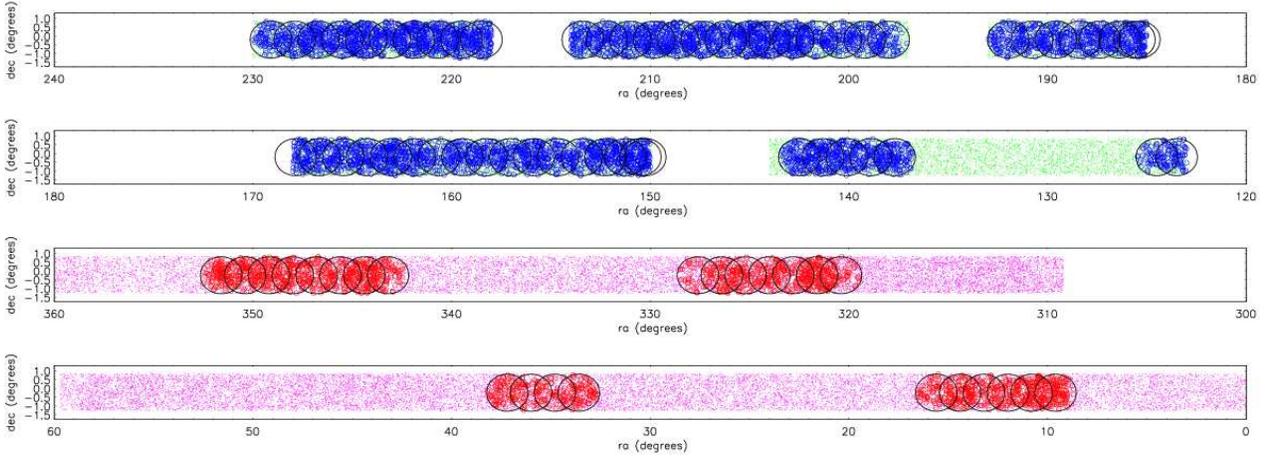}}
\caption{A sky map of the 2SLAQ observations. The upper two panels show the NGC, the lower two panels form the SGC. The black circles represent the 2dF fields observed. Green and pink dots are the NGC and SGC QSO candidates, respectively. The (blue and red) points represent the positions of the (NGC and SGC) spectroscopically confirmed QSOs.}
\label{fig:2SLAQ_strips}
\end{center}
\end{figure*}

The sky regions surveyed by the 2dF instrument consist of two
$2^{\circ}$ -- wide equatorial strips, containing the QSO candidates
observed by SDSS survey. Not all of the full strips were observed, but
rather ``sections'' of them. Fig. \ref{fig:2SLAQ_strips} shows the two
strips, on the NGC and SGC. The NGC photometric candidates are shown in
green and the SGC ones in pink. The blue (red) circles are all the
spectroscopically identified QSOs in the NGC (SGC). The 2dF pointings
are shown as black circles. The ``sections'' in the NGC were indexed
``a, b, c, d, e'' and the one in the SGC ``s''. Each 2dF pointing was
labelled with the index of the region where it fell followed by a
number, which refers to its position along the strip.

The QSO observations were performed simultaneously with those of the
LRGs. 200 2dF fibres were allocated to the LRGs and 200 to the QSO
observations \citep{cannon}. The LRG fibres then link to the 2dF ``red
spectrograph'' and the QSO fibres to the ``blue spectrograph''. Each
block of 10 consecutive fibres along the edge of the 2dF field connects
to a different spectrograph, alternately blue and red. Therefore, the
QSO completeness mask in each 2dF pointing shows a ``dented structure''
along the edge of the field, due to the fact that the fibres are limited
to an angle of $14^{\circ}$ \cite[see, e.g.][]{richards04}. The
probability of a given QSO/LRG candidate being assigned a 2dF fibre
depends on its priority. The assigned priorities of the objects in the
input catalogue (see table \ref{table:prior}) will affect the likelihood
that those objects will be observed. Objects with higher priority will
have a higher likelihood to be assigned a 2dF fibre.

\begin{table}
\centering
\begin{tabular}{ l c } \hline
           Objects                                       & Priority \\ \hline
	   Guide stars                                   & $9$ \\
	   Main sample LRGs, sparsely sampled            & $8$ \\
	   Remaining main sample LRGs                    & $7$ \\
	   $g > 20.5$ QSOs, sparsely sampled             & $6$ \\
	   Remaining $g > 20.5$ QSOs                     & $5$ \\
	   Extra LRGs and high-$z$ QSOs                  & $4$ \\
	   $g < 20.5$ QSOs                               & $3$ \\
	   Previously observed objects with good id      & $1$ \\ \hline
\end{tabular}
\caption[2dF priorities]{2dF priorities. Objects with higher priorities have a higher likelihood of being assigned a 2dF fibre.}
\label{table:prior}
\end{table}

Tables \ref{table:2slaq_ngp} and \ref{table:2slaq_sgp} show the number of QSOs, narrow emission line galaxies (NELGs) and stars that were observed. $Q1$ and $Q2$ refer to the identification quality: $Q1$ are objects with good identification quality and $Q2$ refer to objects with lower identification quality (see section 2.3 of \citealt{scott04} for further details on quality identification flags). Overall, the sky density of QSO candidates in $138.4$ deg$^{-2}$ and that of confirmed QSOs is  $44.7$ deg$^{-2}$.

\begin{table}
\centering
\begin{tabular}{ l c c c } \hline
           ID       & All & Q1 & Q2 \\ \hline
           QSOs     & $6680\ (57.89 \% )$ &    $6482\ (56.17\% )$  &   $198\ ( 1.72\% )$ \\
           NELGs    & $2077\ (18.00\% )$   &    $2043\ (17.71\% )$   &   $34\ ( 0.29\% )$ \\
           stars    & $1829\ (15.85\% ) $   &    $1604\ (13.90\% )$   &   $225\ ( 1.95\% )$ \\
           TOTAL    & $10586\ (92.20\% )$   &   $10129\ (88.15\% ) $   &  $457\ ( 4.05\% )$ \\ \hline

\end{tabular}
\caption[Number of QSOs in the NGC 2SLAQ strip]{Number of QSOs in the NGC 2SLAQ strip.}
\label{table:2slaq_ngp}
\end{table}

\begin{table}
\centering
\begin{tabular}{ l c c c } \hline
           ID       & All & Q1 & Q2 \\ \hline
           QSOs     & $2378 (49.68\% )$  &    $2282 (47.67\% )$  &   $96 ( 2.01\% )$ \\
	   NELGs    & $905 (18.91\% )$   &    $881 (18.40\% )$   &   $24 ( 0.50\% )$ \\
	   stars    & $835 (17.44\% )$   &    $739 (15.44\% )$   &   $96 ( 2.01\% )$ \\
	   TOTAL    & $4118 (86.02\% )$  &    $3902 (81.51\% )$  &   $216 ( 4.51\% )$ \\ \hline
\end{tabular}
\caption[Number of QSOs in the SGC 2SLAQ strip]{Number of QSOs in the SGC 2SLAQ strip.}
\label{table:2slaq_sgp}
\end{table}

\begin{figure*}
\begin{center}
\hskip -6in
\vspace{10.0cm}
\centerline{\epsfysize = 15.0cm, 
\begin{rotate}{-90}
\epsfbox{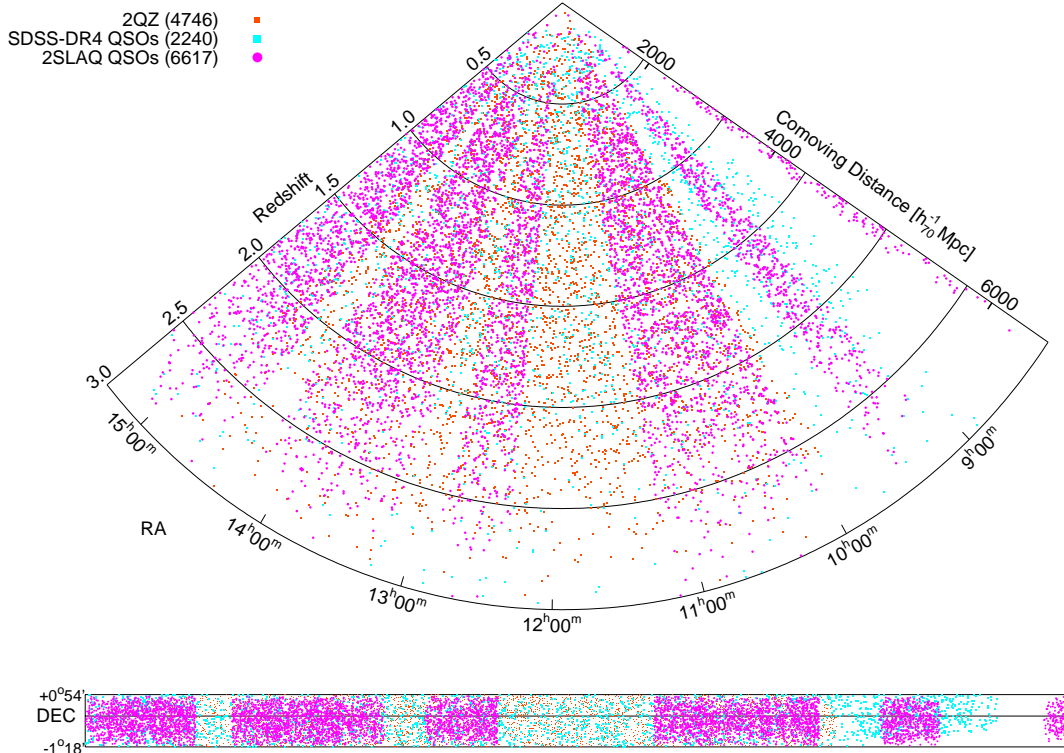}
\end{rotate}}
\caption{This shows a wedge of the sky region covered by the 2SLAQ QSO survey. The other two QSO surveys that have partial coverage in the same area are shown for comparison. The number in brackets gives the total number of QSOs with $z<3$ per survey in this region. The comoving distance is computed assuming a $\Omega_{m}^{0} = 0.3$, $\Omega_{\Lambda}^{0} = 0.7$, $h = 0.7$ cosmology.}
\label{fig:2SLAQ_NGPwedge}
\end{center}
\end{figure*}

As we are observing faint QSOs, we also expect them to have a higher space density than that achieved from other, previous surveys, such as the 2QZ or the SDSS. This is evident from the wedge plot in Fig. \ref{fig:2SLAQ_NGPwedge}, which shows the radial projection of the 2SLAQ NGC strip (in pink). The QSOs observed from the 2QZ and SDSS DR4 \citep{dr4} datasets are also shown in the wedge plot, in blue and cyan squares, respectively.

%\begin{figure*}
%\begin{center}
%\centerline{\epsfxsize = 15.0cm, 
%%\epsfbox{images/2SLAQ/WedgePlots_NGC2.ps}}
%\epsfbox{images/2SLAQ/WedgePlots_NGC_new_rotate_right.ps}}
%\caption{This shows a wedge of the sky region covered by the 2SLAQ QSO survey. The other two QSO surveys that have partial coverage in the same area are shown for comparison. The number in brackets gives the total number of QSOs with $z<3$ per survey in this region. The comoving distance is computed assuming a $\Omega_{m}^{0} = 0.3$, $\Omega_{\Lambda}^{0} = 0.7$, $h = 0.7$ cosmology.}
%\label{fig:2SLAQ_NGPwedge}
%\end{center}
%\end{figure*}

\section{QSO clustering}
\label{sec:clust}

Completeness issues within a 2dF pointing must be taken into account when constructing the angular mask used to generate a random set of points, which is necessary to measure QSO clustering from the 2SLAQ survey. The completeness in each pointing depends on two factors: (i) the {\it coverage} completeness, given by the fraction of QSO candidates that were assigned a 2dF fibre; and (ii) the {\it spectroscopic} completeness, representing the fraction of observed candidates which have good redshift quality. In addition, one needs to calculate the excess probability of finding a QSO in overlapping 2dF pointings. %This excess probability is quantified by determining the ratio between the observed number of QSO candidates and the total number of candidates in the parent catalogue. This {\it fractional completeness} is then used to weight the probability of a QSO being observed  in that region and, as a consequence, it corrects for the different angular completeness in overlapping 2dF pointings. 

\begin{figure}
\begin{center}
\centerline{\epsfxsize = 9.0cm
\epsfbox{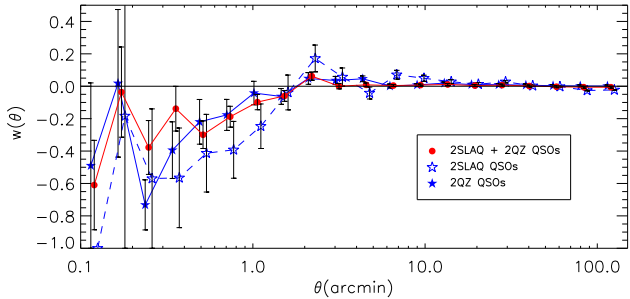}}
\caption{The angular correlation function measured for the 2QZ survey (solid blue stars and solid blue line), the 2SLAQ QSO survey (open blue stars and dashed blue line) and the 2QZ and 2SLAQ QSO surveys combined (red circles and line). The $w(\theta)$ measurements are very similar in both cases and show that the deficit of pairs seen at the smallest scales is not significant at typical QSO-QSO comoving separations. Note that the 2QZ values are offset by a shift of $\Delta \theta = -0.02$ and the 2SLAQ values by a shift of $\Delta \theta = +0.02$.}
\label{fig:w_theta}
\end{center}
\end{figure}

The fact that the 2dF instrument cannot place two fibres any closer than
$\sim 30$ arcsec means that an additional incompleteness can potentially
lead to an artificial deficit of close QSO pairs in 2dF surveys. To make
an approximate correction for these effects, one can measure the angular
correlation function, $w(\theta)$ \citep[e.g.][]{hawk}. Comparing this
to the angular correlation measured in the total input catalogue allows
one to estimate the average deficit of close pairs at small angular
separations. As shown by \cite{scott01}, this deficit is negligible in
the 2QZ sample. In the 2SLAQ sample, however, the deficit of pairs can,
potentially, constitute a bigger bias. This is due to the fact that, in
contrast to what happens in the 2QZ survey, the 2SLAQ QSOs are assigned
a lower observational priority than the main
sample LRGs. Therefore, the QSO-assigned fibres will only be positioned
in areas allowed by the underlying angular distribution of the LRG
fibres. Fig. \ref{fig:w_theta} shows the $w(\theta)$ measurements of the
2QZ+2SLAQ sample and the 2SLAQ and 2QZ samples separately. In order to
better distinguish between the errorbars, the 2SLAQ values are offset by
a shift of $\Delta \theta = +0.02$ and the 2QZ $w(\theta)$ points by a
shift of $-0.02$. To account for the fibre-collision effects in the
clustering  of the 2SLAQ QSOs, we followed the method applied in
previous work to the Two-degree Field Galaxy Redshift (2dFGRS) survey
data \citep{hawk}: the number of QSO pairs at a given separation is
assigned a weight that depends on the QSO's angular separation. Since
the QSO sample spans a wide redshift range, the input catalogue is
expected to show zero correlation at all angular separations $w(\theta)
\sim 0$, $\forall \theta$. In this case, the weight assigned to each QSO
pair using the method of \cite{hawk} is $1/(1+w(\theta))$. The
``imprint'' of the LRG angular distribution on the QSO fibres, due to
these having been assigned a low 2dF priority, is also accounted for:
when generating the random catalogue for the determination of the
correlation functions of the 2SLAQ QSO sample, any random point has a
zero probability of lying closer than $30$ arcsec to any observed LRG.
Although these effects have been considered, 
they have negligible effect on our clustering results.

%In addition to this effect, one should also take into consideration that 2dF fibres cannot overlap or cross.

%At $z\sim 1.5$, this corresponds to $\sim 0.85 \Mpch$, a separation below which the clustering signal will not be significant in our clustering analysis, and most of the existing QSO pairs have very large redshift separations.

\begin{figure}
\begin{center}
\centerline{\epsfxsize = 9.0cm
\epsfbox{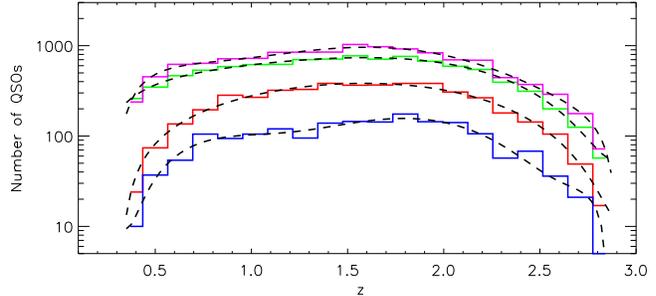}}
\caption{2SLAQ QSO and 2QZ $N(z)$. Red line is the NGC and the blue line
the SGC. The green line represents the 2QZ NGC and the pink line the 2QZ
SGC. Also shown, as dashed lines, are the polynomial fits that were used
to model the radial distribution of the random points.}
\label{fig:N_z2SLAQ}
\end{center}
\end{figure}

Equally as relevant is the radial completeness, which also needs to be
accurately described by the unclustered, or ``random'' distribution.
Fig. \ref{fig:N_z2SLAQ} shows the ($0.3<z<2.9$) redshift distribution of
the 2QZ and 2SLAQ QSOs, in $\Delta z = 0.13$ bins. The red line
represents the 2SLAQ NGC while the blue line the 2SLAQ SGC. The green
and pink lines are the $z$-distributions of the 2QZ NGC and 2QZ SGC
QSOs, respectively. Dashed lines also show the polynomial fits to those
distributions that were used to generate the random distribution.

The 2QZ survey comprises $22416$ (id quality 1) QSOs in the redshift
range $0.3<z<2.9$ ($9982$ in the NGC and $12434$ in the SGC). The 2SLAQ
QSO sample, when imposing faint magnitude cuts ($20.5<g<21.85$) in
addition to these $z$-cuts, comprises a total of $6374$ QSOs ($4574$ in
the NGC and $1800$ in the SGC). The fact that the 2SLAQ $N(z)$ is
steeper, at low-$z$, is possibly due to QSO contamination by host
galaxies, affecting the colour selection of fainter QSOs. The median
redshift of the 2QZ+2SLAQ sample is $<z> = 1.50$.\\

\begin{figure}
\begin{center}
\centerline{\epsfxsize = 9.0cm
\epsfbox{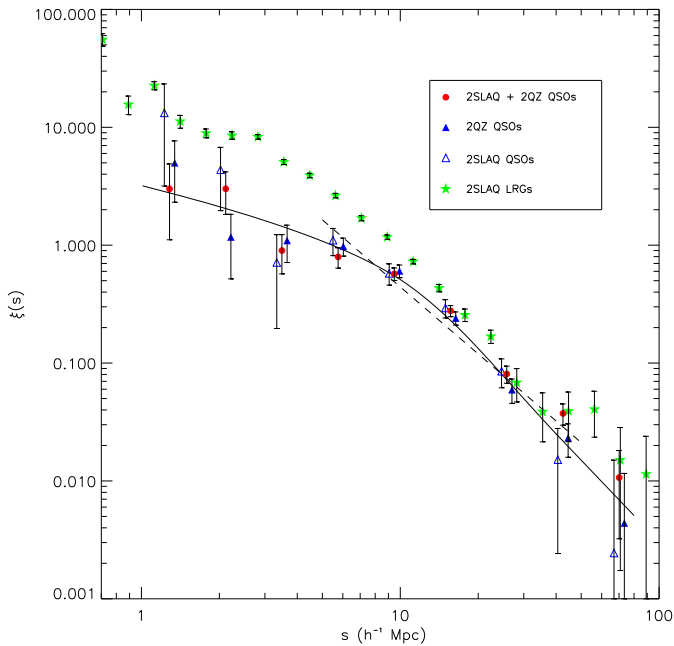}}
\caption{The red circles show the $\xi(s)$ measured from the 2SLAQ and
2QZ samples and the blue triangles the 2QZ results (from da \^{A}ngela
et al., 2005). The $\xi(s)$ measurements are very similar, both in
amplitude and shape. The green stars show the 2SLAQ LRG measurements
\citep{ross}. The dashed and solid lines show two models: the
best fitting 2QZ $5< s< 50 \Mpch$ power law (dashed); and the double
power law $\xi(r)$ model, ``distorted'' by dynamical motions
parameterised by $<w_{z}^{2}>^{1/2}=800 \kms$ and $\beta(z) = 0.32$.
Note that the $\xi(s)$ values from the individual 2SLAQ and 2QZ samples
have been offset by $\log \Delta s$ of $0.02$ and $-0.02$,
respectively.}
\label{fig:2SLAQ_xis}
\end{center}
\end{figure}

After generating a random catalogue we can then combine the new 2SLAQ QSO sample with the 2QZ sample, and compute the QSO clustering by means of correlation functions. We start by estimating $\xi(s)$, the 2-point correlation function measured in $z$-space. This is presented in Fig. \ref{fig:2SLAQ_xis} (filled red circles). The estimator used to measure $\xi(s)$ is the \cite{hamil93} estimator :

\be
\xi(s) = \frac{<\mathrm{DD}(s)><\mathrm{RR}(s)>}{<\mathrm{DR}(s)>^{2}}-1,
\label{equation:hamil_2slaq}
\ee
where $<\mathrm{DD}(s)>$, $<\mathrm{DR}(s)>$, $<\mathrm{RR}(s)>$ are the
mean number of QSO-QSO, QSO-random and random-random pairs at separation
$s$. For comparison, also shown is the previously determined 2QZ
$\xi(s)$ \citep{me2}, the 2SLAQ QSO $\xi(s)$ and also the $\xi(s)$
measurements of the 2SLAQ LRG sample \citep{ross}. To make
the plot clearer, we have offset the 2QZ and 2SLAQ $\xi(s)$ points by a
factor of $0.02$ and $-0.02$, respectively. % and from the SDSS LRG
%sample, \citep{zeha05, eisens}.

Including the 2SLAQ QSO sample does not affect the shape of the
previously measured 2QZ $\xi(s)$. The $\xi(s)$ measured from both
samples, including or not the 2SLAQ QSOs,  are indeed very similar. We
have verified the statistical weight of including the 2SLAQ sample by
comparing the number of QSO-QSO pairs at separations $< 20 \Mpch$, and
verified that the combined 2QZ+2SLAQ sample has $\sim 65\%$ more QSO-QSO
pairs within $20 \Mpch$ than the 2QZ sample alone. This gain also
includes the contribution of the cross pairs between the 2SLAQ and 2QZ
samples, on the NGC strip. The 2SLAQ LRGs have a higher clustering
amplitude than the 2SLAQ QSOs. At smaller scales the two samples also
differ in the shape of their correlation functions. This is
probably due to the different $z$-space distortions that affect the LRGs
and the 2QZ and 2SLAQ QSOs, a contributing factor to which will be the higher 
redshift errors of the QSOs.

Also shown are two different 2QZ $\xi(s)$ models, obtained by
\cite{me2}. The dashed line is the best fitting 2QZ power-law model, in
the range $5< s< 50 \Mpch$ ($\xi(s) = (s/6.50)^{-1.89}$), and the solid
line is the $\xi(s)$ model obtained from convolving a double power-law
$\xi(r)$ model (Eq. \ref{equation:xir2pl2qz}) with the $z$-space
distortions parameterised by $<~w_{z}^{2}>^{1/2}=800\kms$ and $\beta(z)
= 0.32$.

\be
\xi(r) = \left\{
\begin {array}{l}
(r/6.00)^{-1.45},\ \ r<10 \Mpch\\
\noalign{\medskip}
(r/7.25)^{-2.30},\ \ r>10 \Mpch
\end {array}
\right.
\label{equation:xir2pl2qz}
\ee

It can be seen that the model is still a good description of the joint
QSO $\xi(s)$ measurements, indicating that the 2SLAQ QSOs should have a
similar real-space clustering and be subjected to the same dynamical
distortions as the 2QZ QSOs. The fitting of these models does not take
into consideration the correlations between the errors at different
separations. However, \cite{thesis} showed that taking into account the
full covariance matrix  when fitting the 2QZ $\xi(s)$
does not affect the $s_{0}$ and $\gamma$ values by more than $1 \sigma$.

\begin{figure}
\begin{center}
\centerline{\epsfxsize = 9.0cm
\epsfbox{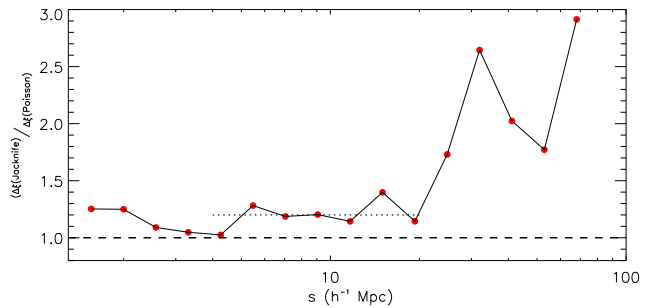}}
\caption{Red circles and solid line show the ratio jacknife and Poisson $\xi(s)$ errors. Poisson errors seem to under-predict the uncertainty in $\xi(s)$ at all scales, and considerably at the largest scales. At intermediate, $4\lsim s \lsim 20 \Mpch$ scales, the ratio of the two error estimates is approximately constant and $\sim 1.25$ (dotted line).}
\label{fig:errors}
\end{center}
\end{figure}

The errors shown in Fig. \ref{fig:2SLAQ_xis} are ``jacknife'' estimates, estimated by splitting the 2QZ+2SLAQ sample in $16$ subsamples. We compared the jacknife and Poisson error estimates in our $\xi(s)$ computation. The Poisson error estimates should, in principle, provide a fair description of the uncertainty for the 2QZ QSO clustering measurements \citep{fiona00,me2}. Here we test this hypothesis for the new sample containing the 2QZ and 2SLAQ QSOs. We divide up the overall 2QZ+2SLAQ dataset into $16$ subsamples and compute $\xi(s)$ in the overall set minus each of the $16$ subsamples in turn\footnote{This $\xi(s)$ computation was performed using the $k$d-tree algorithm of \cite{kdtree}.}. The $16$ measurements of $\xi(s)$ are then combined as follows, in order to obtain the jacknife error \citep[e.g.][]{adamnew}:

\be
\sigma_{\mathrm{jacknife}} = \sqrt{\sum_{i=1}^{N}\frac{\mathrm{DR}_{i}(s)}{\mathrm{DR}_{tot}(s)}(\xi_{i}(s)-\xi_{tot}(s))^{2}}
\label{equation:jack_err_2slaq}
\ee
where $N$ is the total number of subsamples ($16$, in this case); the subscript $i$ refers to the whole dataset minus subsample $i$; and $tot$ refers to the whole 2QZ+2SLAQ QSO sample. The ``$\mathrm{DR}$ ratio'' accounts for the fact that the subsamples may not necessarily contain exactly the same number of QSOs. Fig \ref{fig:errors} shows the ratio between the jacknife and the Poisson errors. It can be seen that, on all scales, Poisson errors underestimate the uncertainty on the clustering measurements, especially at the largest scales. On scales $2\lsim s \lsim 4 \Mpch$, the two estimates are quite similar, but on $4\lsim s \lsim 20 \Mpch$ scales, where most of the clustering signal is obtained, the jacknife errors are, on average, $1.25$ times bigger than Poisson errors (dotted line). At larger scales, where there are fewer QSO independent pairs, the Poisson estimates largely under-predict the true error estimate as has been previously discussed, \citep[e.g.][]{sboyle,adamnew}.\\

\begin{figure}
\begin{center}
\centerline{\epsfxsize = 9.0cm
\epsfbox{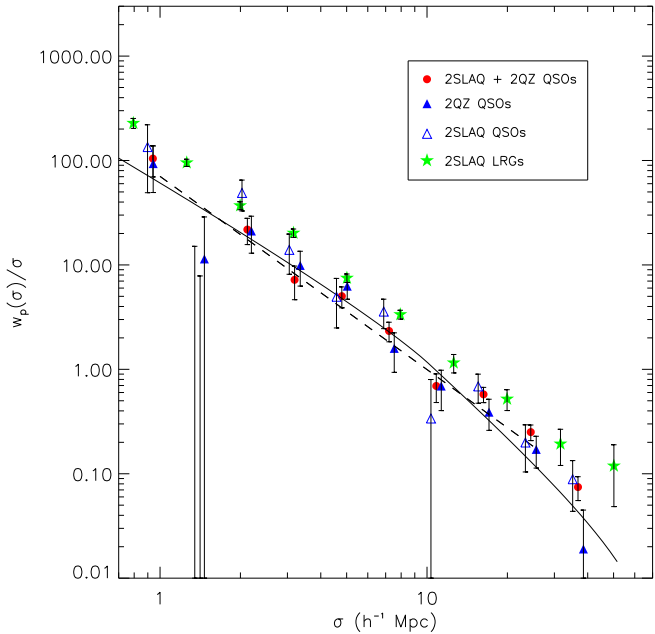}}
\caption{The red circles are the $w_{p}(\sigma)/\sigma$ measurements for the 2QZ+2SLAQ sample. These are very similar to those of the 2QZ sample alone (blue triangles; da \^{A}ngela et al., 2005). The green stars represent the higher clustered  LRG sample from the 2SLAQ survey \citep{ross}. The models shown represent the projection of a single (dashed line) and a double (solid line) power law models. Note that the $w_{p}(\sigma)/\sigma$ values from the individual 2QZ and 2SLAQ samples have been offset by $\log \Delta \sigma$ of $0.02$ and $-0.02$, respectively.}
\label{fig:2SLAQ_wp}
\end{center}
\end{figure}

Fig. \ref{fig:2SLAQ_wp} shows the projected correlation function measured from the 2QZ+2SLAQ sample (red circles). This is very similar to the previous 2QZ measurement (\citealt{me2}, blue triangles, offset by a factor of $\log \Delta \sigma = 0.02$). The open blue triangles represent the $w_{p}(\sigma)/\sigma$ values for the 2SLAQ ensemble alone (offset by a factor of $\log \Delta \sigma = -0.02$) and the green stars represent the more strongly clustered 2SLAQ LRGs \citep{ross}. The solid line is the $\sigma$-projection of the double power-law $\xi(r)$ model which was found to be a good description of the 2QZ $\xi(r)$. The relation between $w_{p}(\sigma)$ and $\xi(r)$ is given by:

\be
w_{p}(\sigma) = 2\int_{\sigma}^{\infty}\frac{r\xi(r)}{\sqrt{r^{2}-\sigma^{2}}}dr
\label{equation:wp_sig_xi_r}
\ee

The dashed line corresponds to the projection of a power law $\xi(r)$ model, given by $\xi(r) = (r/4.96)^{-1.85}$.

\begin{figure}
\begin{center}
\centerline{\epsfxsize = 9.0cm
\epsfbox{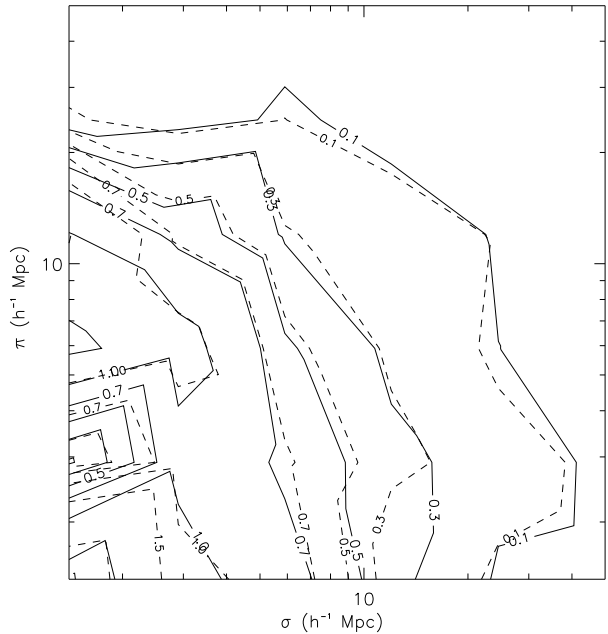}}
\caption{ $\xi(\sigma,\pi)$ measured for the 2QZ+2SLAQ sample (solid contours) and for the 2QZ sample alone (dashed contours). The two measurements are very similar.}
\label{fig:2SLAQ_xisp}
\end{center}
\end{figure}

The fact that the 2SLAQ survey targeted faint QSOs is not only an advantage for studies of the luminosity-dependence of QSO clustering, but also for $z$-space distortion analyses. The higher spatial density of the combined QSO sample should, in principle, improve our statistics when studying $z$-space distortions, and, in particular, the estimation of $\Omega_{m}^{0}$ and $\beta(z)$ from dynamical and geometrical $\xi(\sigma,\pi)$ distortions. The $\xi(\sigma,\pi)$ measured from the whole QSO sample is shown in Fig. \ref{fig:2SLAQ_xisp} (solid contours). The dashed lines refer to the 2QZ measurement.

\section{Parameter constraints from redshift-space distortions}
\label{sec:zdist}

There are basically two mechanisms leading to dynamical $z$-space distortions. As structures grow through gravity, the infall of objects to higher-density regions contributes to the measured redshifts. If these are assumed to be solely due to the Hubble flow, then the large-scale distribution will appear flatter, or thinner, along the line of sight, thus ``distorting'' the clustering signal. At smaller scales, the random peculiar motions of the objects will also contribute to the measured redshifts, and hence distort the measured clustering signal for close pairs of objects. If the distribution of distant objects has, on average,  a spherically symmetric clustering pattern in real space, but large velocity dispersion, then the clustering signal measured in $z$-space will be smeared along the line-of-sight. These features are often referred to as ``fingers-of-God'', and are commonly seen as elongated structures in radial wedge plots of distant galaxy surveys, such as the 2dFGRS. 

Peculiar velocities are not the only effect leading to anisotropies in the clustering pattern. As shown by \cite{ap}, if one assumes a cosmology different from the true, underlying cosmology of the Universe to convert redshifts into distances, the effect on separations along the line of sight differs from that affecting the separation in the angular coordinate. As a consequence, the clustering signal might appear elongated (or squashed) in the redshift direction. As shown by those authors these {\it geometric distortions} can be a powerful cosmological test, namely to determine $\Omega_{m}^{0}$. 

Due to their significance at high-$z$, these potential geometric distortions have been used to constrain cosmological parameters using QSO catalogues \citep[e.g.][]{phil04}; 21 cm maps of the epoch of reionisation \citep{nusser} or the Lyman $\alpha$ forest \citep{becker}. 

However, and as discussed in detail in \cite{bph}, it is sometimes not trivial to disentangle the effects of geometric distortions from those caused by peculiar velocities. If both the infall parameter $\beta$ and cosmological parameters as $\Omega_{m}^{0}$ or $\Omega_{\Lambda}^{0}$ are left as free variables, we expect to see a degeneracy between the anisotropies caused by the large scale infall and the geometric distortions. Those authors define a ``flattening factor'', which determines, as a function of redshift and cosmology, the level of asymmetry expected to be seen as a result of geometric distortions, and found that its value is degenerate with that of $\beta$.

The fitting of the dynamical and geometrical distortions in $\xi(\sigma,\pi)$ is described in detail in section 7.7 of \cite{me2}. In summary: 

{\bf 1)} for a given value of $\beta(z)$, a $\xi(\sigma,\pi)$ model is generated in a chosen test cosmology, through \citep{msuto}:

\begin{eqnarray}
\xi(\sigma,\pi) &= &\left(1+\frac{2}{3}\beta(z)+\frac{1}{5}\beta(z)^{2}\right)\xi_{0}(r)P_{0}(\mu)\\
                &- &\left(\frac{4}{3}\beta(z)+\frac{4}{7}\beta(z)^{2}\right)\xi_{2}(r)P_{2}(\mu)\\
                &+ &\frac{8}{35}\beta(z)^{2}\xi_{4}(r)P_{4}(\mu),
\label{equation:hamilton2}
\end{eqnarray}
where $\mu$ is now the cosine of the angle between $r$ and $\pi$ and $P_{l}(\mu)$ are the Legendre polynomials of order $l$. $\xi_{0}(r)$, $\xi_{2}(r)$ and $\xi_{4}(r)$ are the moments of order $0$, $2$ and $4$ of the linear $\xi(r)$ and their form depends on the $\xi(r)$ model adopted. In general, they are given by \citep{msuto}:

\be
\xi_{2l}(r) = \frac{(-1)^{l}}{r^{2l+1}}\left(\int_{0}^{r}xdx\right)^{l}x^{2l}\left(\frac{d}{dx}\frac{1}{x}\right)^{l}x\xi(x)
\label{equation:hamilt_xigen}
\ee

{\bf 2)} The $\xi(\sigma,\pi)$ model is then convolved with the pairwise peculiar velocity distribution to include the small scale $z$-space effects due to the random motions of the QSOs:
\be
 \xi(\sigma,\pi) = \int_{-\infty}^{\infty} \xi'(\sigma,\pi-w_{z}(1+z)/H(z)) f(w_{z}) dw_{z},
\label{equation:final_xisp}
\ee
where the pairwise velocity distribution $f(w_{z})$ can be well described by a Gaussian \citep{ratcliffe}:

\be
f(w_{z}) = \frac{1}{\sqrt{2\pi}<w_{z}^2>^{1/2}}\exp\left(-\frac{1}{2}\frac{|w_{z}|^{2}}{<w_{z}^2>}\right)
\label{equation:disp_vel}
\ee

{\bf 3)} Then, the separations $\sigma$ and $\pi$ are scaled to the same cosmology that was assumed to measure the actual data. The final model for $\xi(\sigma,\pi)$ is then compared to the data. 

The relation between the separations $\sigma$ and $\pi$ in the test and
assumed cosmologies (referred to by the subscripts $t$ and $a$,
respectively) is the following \citep{bph}:

\be
\sigma_{t} = f_{\perp} \sigma_{a} = \frac{B_{t}}{B_{a}} \sigma_{a}
\label{equation:sig_cosmol}
\ee
\be
\pi_{t} = f_{\parallel} \pi_{a} = \frac{A_{t}}{A_{a}} \pi_{a}
\label{equation:pi_cosmol}
\ee
where $A$ and $B$ are defined as (for spatially flat cosmologies):

\be
A = \frac{c}{H_{0}}\frac{1}{\sqrt{\Omega_{\Lambda}^{0}+\Omega_{m}^{0}(1+z)^{3}}}
\label{equation:A_cosmol}
\ee

\be
B = \frac{c}{H_{0}}\int_{0}^{z}\frac{dz'}{\sqrt{\Omega_{\Lambda}^{0}+\Omega_{m}^{0}(1+z')^{3}}}.
\label{equation:B_cosmol}
\ee

In the linear regime, the correlation function in the assumed cosmology will be the same as the correlation function in the test cosmology, given that the separations are scaled appropriately. i.e.:
\be
\xi_{t}(\sigma_{t},\pi_{t}) = \xi_{a}(\sigma_{a},\pi_{a}).
\label{equation:xisp_cosmol}
\ee

{\bf 4)} This method is repeated for different test cosmologies and values of $\beta(z)$. \\

Given the similarities between the $\xi(\sigma,\pi)$ contours in Fig.
\ref{fig:2SLAQ_xisp}, in addition to very similar $\xi(s)$ and
$w_{p}(\sigma)$ measurements, we would not expect the constraints put on
$\beta(z)$ and $\Omega_{m}^{0}$ from the 2QZ+2SLAQ dynamical distortions
to differ from those obtained from the 2QZ sample alone \citep{me2},
assuming that all the underlying assumptions remain the same (e.g.,
$\xi(r)$ shape and amplitude, velocity dispersion, scale-independent
bias). We now repeat the method adopted for fitting the 2QZ dynamical
and geometrical distortions, but also utilising the new 2SLAQ ensemble.
The question now arises if the same $\xi(r)$ model should be assumed, or
if the velocity dispersion of the QSOs should still be fixed at $800
\kms$. It can be seen in Figs. \ref{fig:2SLAQ_xis} and
\ref{fig:2SLAQ_wp} that the 2QZ double power-law $\xi(r)$ model is still
a good description of both $\xi(s)$ and $w_{p}(\sigma)$ measurements for
the combined sample. As the 2QZ and 2SLAQ samples have similar $N(z)$ we
would not expect to see clustering differences between them due to
redshift evolution. Any potential clustering difference between both
sets would be due to the different luminosity of the samples. However,
as suggested by both observations
\citep[e.g.][]{scottnew,adel06,porc_peder,myers06b}, and simulations
\citep{lidz06,hopkins5} and, more importantly, as we shall see later in
this paper, QSO clustering is very weakly luminosity-dependent. We
therefore assume the same double power-law $\xi(r)$ prescription as used
for the 2QZ sample. We also assume the same velocity dispersion as for
the 2QZ sample alone. It is not unlikely that the 2SLAQ QSOs would have,
on average, a different velocity dispersion. As pointed out by
\cite{berlind03}, \cite{yosh03}, or \cite{tinker06}, galaxies can be a
biased tracer of the dark matter velocity distribution, just as they are
of the dark matter spatial distribution. However, as found for the
2dFGRS galaxies and predicted by HOD (Halo occupation distribution)
models \citep{tinker06}, the expected difference for $M_{b_{J}}\lsim
-20$ is not significant. In addition, as most of the $z$-error is due to
measurement error rather than intrinsic velocity dispersion
\citep{scottnew}, we chose to continue assuming $<w_{z}^{2}>^{1/2} = 800
\kms$.

\begin{figure}
\begin{center}
\centerline{\epsfxsize = 9.0cm
\epsfbox{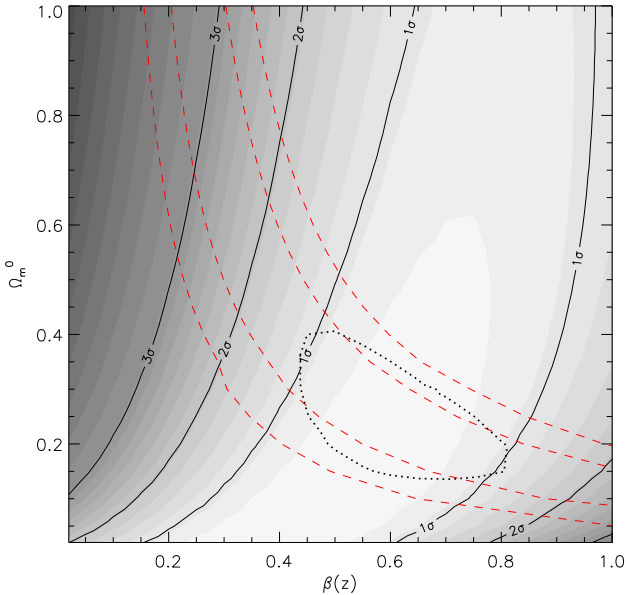}}
\caption{Confidence levels in the $[\Omega_{m}^{0}, \beta(z)]$ plane
obtained from fitting the 2QZ and 2SLAQ $\xi(\sigma,\pi)$ $z$-space
distortions (solid lines and shaded contours). Dashed lines show the $1\
\sigma$ and $2\ \sigma$ constraints from linear theory evolution. The
dotted contour is the $1\ \sigma$ joint confidence level.}
\label{fig:2SLAQ_ombeta}
\end{center}
\end{figure}

The fit to the distortions in $\xi(\sigma,\pi)$ was performed with the
same assumptions and over the same range of scales as in the previous
2QZ analysis. The result is shown in Fig. \ref{fig:2SLAQ_ombeta}. As
expected, the contours are indeed tighter than the ones obtained when
fitting only the 2QZ $\xi(\sigma,\pi)$. This is due to the increased
number of pairs, not only from the 2SLAQ sample alone but also from the
cross-pairs in the NGC between the two ensembles, as they probe
overlapping volumes. Also shown are the $1 \sigma$ and $2 \sigma$
confidence levels predicted from clustering evolution and linear theory
of density perturbations (dashed lines). The dotted line is, as usual,
the $1 \sigma$ joint confidence levels from both constraints. The best
fitting values are $\Omega_{m}^{0} = 0.25_{-0.07}^{+0.09}$, $\beta(z) =
0.60_{-0.11}^{+0.14}$, corresponding to a $\chi^{2}_{min} = 1.02$ (12
d.o.f.). Although these results favour a somewhat higher value of
$\beta$ than the previous 2QZ only $\xi(\sigma,\pi)$ constraint, both
obtained results are self-consistent, within the associated errors. We
should point out that the size of the error bars does not take into
account any potential correlation between $\xi(\sigma,\pi)$ bins but
this is expected to be small. Finally the above derived values for
$\Omega_{m}^{0}$ and $\beta$ imply  a value of the  QSO bias of
$b(z=1.4)=1.5\pm0.2$ which is slightly lower than, but not inconsistent
with, the value of $b(z=1.4)\approx2$ derived below from purely the QSO
clustering amplitude.

\section{Luminosity-redshift degeneracy}
\label{sec:Lz}

A few recent works have looked at the evolution of QSO clustering
\citep[e.g.][]{scottnew,porciani,myers06}. These suggest an increase of
QSO clustering amplitude with redshift, a trend which is more
significant at $z \gsim 1.6$. This evolution contrasts with that
expected from a long-lived QSO population model, or linear theory
predictions, which generally predict a decrease of clustering amplitude
with increasing redshift \citep{scott01,scottnew}. The range of
magnitudes covered by the QSO surveys used in these studies has not
fully permitted the study of the luminosity dependence of QSO
clustering. However, the combination of the 2QZ and 2SLAQ samples
probably sees its greatest scientific contribution precisely in the
range of luminosities it probes and for the first time allows a more
rigorous determination of the QSO clustering dependence on luminosity.
\cite{croom02} have used the 2QZ sample alone for this purpose. Their
results suggest that additional, fainter data, such as
those obtained for 2SLAQ, are essential to pursue this goal.\\

To estimate the $b_{J}$ band absolute magnitude, $M_{b_{J}}$, we compute:

\be
M_{b_{J}}(z) = b_{J}-K_{b_{J}}(z)-A_{b_{J}}+5-5\log(d),
\label{equation:MbJ_mJ}
\ee
where $b_{J}$ is the apparent magnitude, $K_{b_{J}}$ the k-correction in the $b_{J}$ magnitude, $A_{b_{J}}$ the dust correction and $d$ the luminosity distance that corresponds to the redshift $z$, measured in parsecs. The value of the k-correction was taken from \cite{cristiani_vio}. The galactic dust correction, $A_{b_{J}}$ is determined through: $A_{b_{J}} = 4.035 E(B-V)$ \citep{dust}.

The above formula is used to determine the absolute magnitude of the 2QZ
QSOs. To include the dust correction when determining the absolute
magnitude of the 2SLAQ QSOs, one subtracts the $g$ magnitude galactic
extinction ($g_{red}$) at the QSO's coordinates from the observed
apparent magnitude ($g$): $g' = g-g_{red}$, where $g'$ is the
dust-corrected $g$-band QSO magnitude. The other subtlety in combining
the two QSO samples is accounting for the relation between the observed
$b_{J}$ and $g$ magnitudes. However, this becomes quite simple as the
transmissivity curves of the filters have a significant overlap and the
same zero-point. Thus, we can treat these bands as being equivalent
\citep{richards04}. Hereafter, and for the sake of simplicity, we shall
refer to the QSO absolute magnitudes for both samples as if they had
been measured in the $b_{J}$ band, and represent both of them as
$M_{b_{J}}$. Therefore, the 2SLAQ QSOs' absolute magnitude is determined
by:

\be
M_{b_{J}}(z) \approx g'-K_{b_{J}}(z)+5-5\log(d),
\label{equation:Mg_mg}
\ee
where $g'$ already includes the dust correction in the $g$ band.

\begin{figure}
\begin{center}
\centerline{\epsfxsize = 9.0cm
\epsfbox{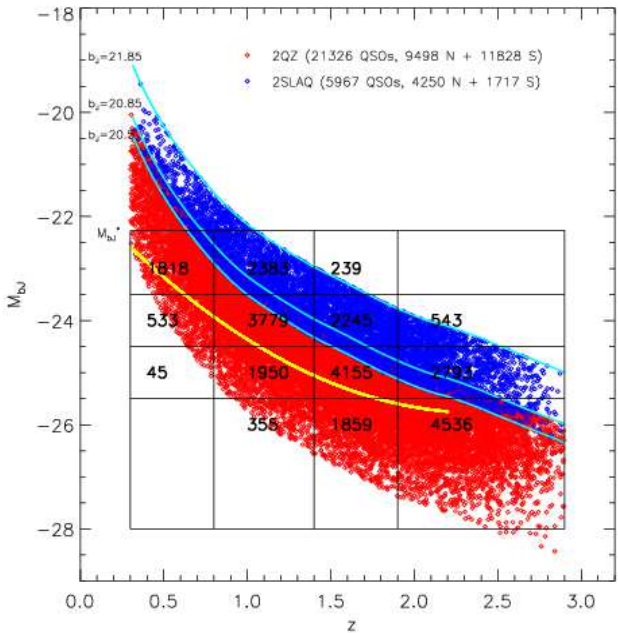}}
\caption{Magnitude and redshift bins adopted for 2QZ and 2SLAQ QSOs. The numbers in each division of the ``grid'' are the numbers of 2QZ and 2SLAQ QSOs in the specific $z$ and $M_{b_{J}}$ ranges.}
\label{fig:Mz2}
\end{center}
\end{figure}

Fig. \ref{fig:Mz2} shows how the 2QZ and 2SLAQ are distributed in the $[M_{b_{J}},z]$ plane. The 2QZ QSOs are shown in red and the 2SLAQ in blue. The cyan lines represent the adopted 2QZ $b_{J}<20.85$ and 2SLAQ $20.5<g<21.85$ magnitude cuts. The QSO samples span the $z$-range $0.3<z<2.9$. The yellow line shows how $M_{b_{J}}^{*}$ changes with $z$. We adopted a second-order polynomial model to determine $M_{b_{J}}^{*}(z)$ \citep{boyle,scott04,richards04}:

\be
M_{b_{J}}^{*}(z) = M_{b_{J}}^{*}(0)-2.5(k_{1}z-k_{2}z^{2})
\label{equation:Mg_star}
\ee

We adopt the values obtained by \cite{scott04}: $M_{b_{J}}^{*}(0) =
-21.61$, $k_{1} = 1.39$, $k_{2}= -0.29$. \cite{richards04} showed that
the parameterisation of the $M_{b_{J}}^{*}(z)$ model is only marginally
affected by including or not the 2SLAQ QSOs. The yellow line in Fig.
\ref{fig:Mz2} only extends to $z=2.2$ given the fitting range used in
this parameterisation.

The flux-limited nature of these two surveys is evident in this plot.
More luminous QSOs lie at higher redshifts while fainter ones have lower
redshifts. This means that, unless we probe a wide window in
magnitude-space with our QSO surveys, it will be intrinsically hard to
determine how QSO physical properties change with luminosity, for a
fixed redshift. By combining the 2SLAQ and 2QZ samples we are widening
the magnitude window and hence making it possible to determine the
dependence of QSO clustering on luminosity, free of any evolutionary
effects.

Fig. \ref{fig:Mz2} also shows how, using the two surveys together, we
can look at a specific redshift range and determine the QSO clustering
in different magnitude samples. This ``vertical approach'' to the
$[M_{b_{J}},z]$ distribution is possibly more physically justifiable
than simply analysing QSO clustering dependence on redshift or apparent
magnitude. Tests of models where comparisons at fixed luminosity are
required certainly need as full coverage as possible of the
luminosity-redshift plane. Indeed, the long-lived QSO model has been
easiest to test in previous samples, since comparing intrinsically low
luminosity QSOs at low redshift with high luminosity QSOs at high
redshift makes more sense in a PLE model where the two are hypothesised
to be directly related. These results have been used to argue against a
long-lived model for QSOs with the 2QZ results of $\xi_{20}$ (see equn.
\ref{equation:xi20} below) appearing to rise, if anything, rather than
fall with increasing redshift \citep{scott04}. However, the  low
redshift ($z\approx0.02$) IRAS selected Seyfert 1 and 2 results of
\cite{georg_shanks} give $\xi_{20}=0.52\pm0.13$ in good agreement
with the low redshift SDSS AGN results of \cite{wake04} which give
$\xi_{20}=0.48\pm0.03$ and adding these relatively high amplitude points
to Fig. 21a of \cite{scott04} may make the clustering  case against
the long-lived model less strong. We note also that the low value of
$\xi_{20}=0.22\pm0.08$ measured for the cross-correlation of low
luminosity QSOs with Lyman-break galaxies   at z=2.5 by \cite{adel06}
also goes against the trend for higher clustering amplitudes for higher
redshift QSOs. But whatever the hypothesis for QSO lifetime, the
extended luminosity range of  the 2SLAQ sample means that we can now 
test the generic prediction of these `high peaks' bias models for higher
clustering amplitudes for more luminous, rare QSOs at fixed redshift.

\section{Clustering as a function of magnitude and redshift}
\label{sec:Lz_xi}

\begin{figure}
\begin{center}
\centerline{\epsfxsize = 9.0cm
\epsfbox{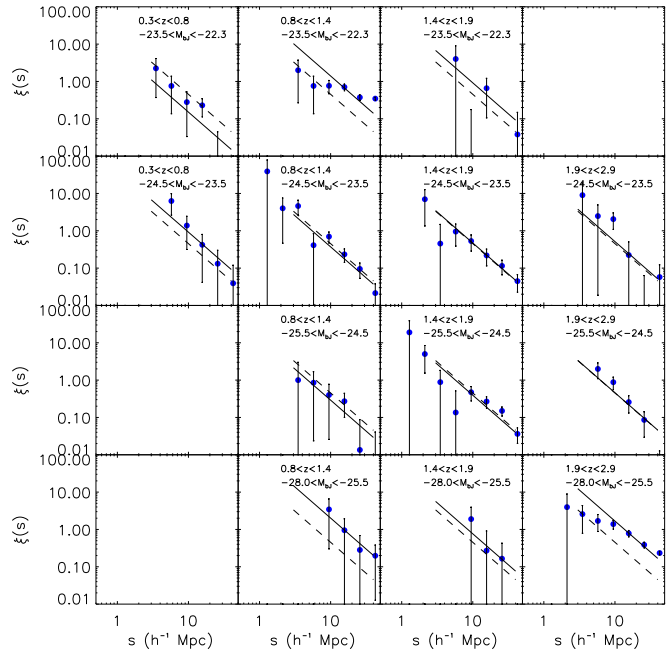}}
\caption{QSO $\xi(s)$ measured in different magnitude and redshift bins. The order of the panels is the same as that of the $[M_{b_{J}},z]$ intervals in Fig. \ref{fig:Mz2}. The dashed line shows the best fitting power-law to the $\xi(s)$ of the full sample. The solid line is the $\xi(s)$ power-law fit to the data in each individual panel.}
\label{fig:xi_s4lum4z}
\end{center}
\end{figure}

Dividing up the QSO samples into magnitude and redshift bins
significantly increases the error on our clustering measurements, simply
due to the much smaller number of objects in each bin compared to the
total number of QSOs (numbers in Fig. \ref{fig:Mz2}). This is also
evident in Fig. \ref{fig:xi_s4lum4z}, where we plot the $\xi(s)$
measurements in each of the panels in Fig. \ref{fig:Mz2}. The dashed
line shows the best fitting power-law model to the overall 2QZ+2SLAQ
sample, over the $3 < s < 50\ \Mpch$ range ($\xi(s) =
(s/6.20)^{-1.66}$). The solid lines are the best power-law models to
each individual $[M_{b_{J}},z]$ interval, fixing the $\xi(s)$ slope to
$\gamma = 1.66$ and performing a $\chi^{2}$ fit to determine the
amplitude. The order of the panels in Fig. \ref{fig:xi_s4lum4z} is the
same as in the panels presented in the $[M_{b_{J}},z]$ plane in Fig.
\ref{fig:Mz2}.

By visually comparing the dashed and solid lines, we observe no
dependence of QSO clustering on luminosity nor redshift. However, the
size of the errorbars motivates the further use of more statistically
robust tools. We therefore use the integrated correlation function up to
$20 \Mpch$ in order to quantify the clustering amplitude in each
magnitude-$z$ bin. This quantity is then normalised to the volume
contained in a $20 \Mpch$ sphere:

\be
\xi_{20} = \frac{3}{20^3}\int_{0}^{20}\xi(s)s^2ds
\label{equation:xi20}
\ee

The choice of using $20 \Mpch$ as the radius of the spheres to compute
the averaged correlation function is due to the fact that this is a
large enough scale for linear theory to be applied and, as shown by
\cite{scottnew}, small-scale $z$-space distortions do not significantly
affect the clustering measurements, when averaged over this range of
scales. In addition, and as seen in Fig. \ref{fig:errors}, we can
estimate the uncertainty through computing Poisson errors, and scale
this by a factor of $1.25$. This estimate should provide a fair
description of the uncertainty on the correlation function measurements,
and significantly reduce the computing time.

\begin{figure}
\begin{center}
\centerline{\epsfxsize = 9.0cm
\epsfbox{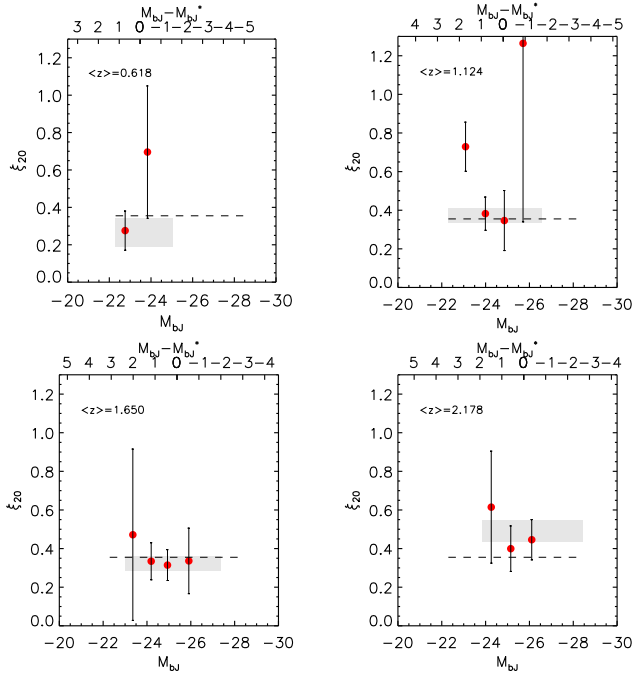}}
\caption{The four panels represent the $\xi_{20}$ measurements in different redshift bins. The median redshift of each $z$-interval is indicated in the top left of each graph. The top horizontal axis shows the absolute magnitude difference, relative to $M_{b_{J}}^{*}(<z>)$. The red circles are the $\xi_{20}$ measurements in different absolute magnitude bins, and are centred on the median values of each bin. The shaded area is the $1 \sigma$ $\xi_{20}$ interval for all the QSOs in that specific redshift interval. The horizontal length of the shaded area represents the range of $M_{b_{J}}$ values of QSOs in that redshift interval.}
\label{fig:xi20_4lum4z}
\end{center}
\end{figure}

We computed $\xi_{20}$ using the Hamilton estimator in each of the bins
shown in Fig. \ref{fig:Mz2}. The results for each redshift slice are
shown in the four panels in Fig. \ref{fig:xi20_4lum4z}. Red circles show
the measurements in each magnitude bin. The shaded grey area shows the
$1 \sigma$ $\xi_{20}$ measurement for QSOs of all luminosities in that
specific redshift slice and its length indicates the total range of
magnitudes included. The dashed line represents the average value of
$\xi_{20}$, for all redshift and magnitude ranges. It should be pointed
out that the bin sizes were chosen in such a way that the precision of
the clustering measurements was maximised, and therefore the
distribution of QSOs in a given $z$-slice is not constant for all
magnitudes. Thus, we do not expect our $\xi_{20}$ measurements to be
equidistant along the horizontal axis, as these are centred on the
median values in $M_{b_{J}}$ of each bin. The top axis indicates the
magnitude difference with respect to $M_{b_{J}}^{*}(<z>)$, at the median
redshift of that specific ``$z$ - slice''. The ``rising'' of the grey
area as we move to higher redshifts is consistent with the results of
\cite{scottnew}, who also found an increase of clustering amplitude with
redshift, for the 2QZ QSOs.

The number of QSOs in each $M_{b_{J}} - z$ bin, indicated in Fig.
\ref{fig:Mz2}, is now reflected in the sizes of the $\xi_{20}$ error
bars. In the first, lower-$z$ panel, for instance, the $M_{b_{J}} - z$
bin with only $533$ QSOs corresponds to the $\xi_{20}$ measurement with
the largest error bar. The two intermediate $z$-slices are the ones
where most of the gain of the 2SLAQ is observed and the ones with
highest statistical value. Our results are in  agreement with the
hypothesis of a luminosity-independent clustering ($\chi^{2}_{red} =
1.16$, over 12 d.o.f.).  The hypothesis of QSO clustering being constant
with redshift and luminosity is not supported by the data
($\chi^{2}_{red} = 2.50$).

\section{Bias and halo masses}
\label{sec:Lz_bias}

The $\xi_{20}$ vs. $M_{b_{J}}$ results motivate the analysis of the dependence of bias on luminosity and redshift. \cite{scottnew} investigated the redshift evolution of QSO bias, using the 2QZ survey data. They found that the QSO bias does evolve very strongly with redshift; as the mass clustering amplitude decreases with increasing redshift, the slight upward trend observed in the 2QZ $\xi_{20}$ reveals a strong increase of bias with $z$. %We here study this bias evolution by investigating the effects of QSO luminosity in the measured value of $b(z)$.

Under the assumption of a scale-independent bias, the bias can be
obtained through \citep[e.g.][]{peebles}:

\be
b = \sqrt{\frac{\xi_{Q}(r)}{\xi_{\rho}(r)}}\approx \sqrt{\frac{\xi_{Q}(r,20)}{\xi_{\rho}(r,20)}},
\label{equation:biasxi20}
\ee
where $\xi_{Q}(r,20)$ and $\xi_{\rho}(r,20)$ represent the QSO and
matter real-space correlation functions, respectively, averaged in $20
\Mpch$ spheres. The $z$-space and real-space correlation functions can
be given by \citep{kaiser}:

\be
\xi_{Q}(s,20) = \left(1+\frac{2}{3}\beta+\frac{1}{5}\beta^{2}\right)\xi_{Q}(r,20)
\label{equation:biasxi202}
\ee

Combining  both equations and taking into account that $\beta =
\Omega_{m}^{0.6}/b$ leaves us with a quadratic equation in $b$. Solving
it (see \citealt{scottnew}) leads to:

\be
b(z)=\sqrt{\frac{\xi_{Q}(s,20)}{\xi_{\rho}(r,20)}-\frac{4\Omega_{m}^{1.2}(z)}{45}}-\frac{\Omega_{m}^{0.6}(z)}{3}
\label{equation:biasxi203}
\ee

\begin{figure}
\begin{center}
\centerline{\epsfxsize = 9.0cm
\epsfbox{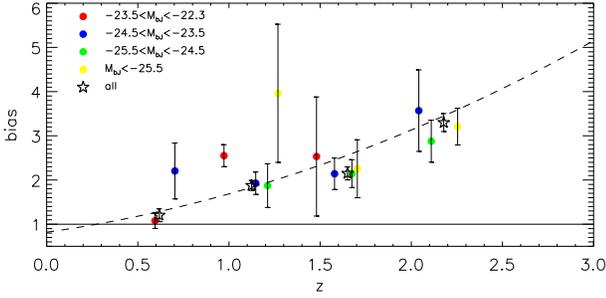}}
\caption{Bias evolution for different luminosity QSOs. The different colours refer to different absolute magnitude bins. The stars are the result for all the QSOs in each specific redshift bin. The dashed line is the empirical model of Croom et al. (2005). Each point is represented in the median redshift of all the QSOs in the specific $M_{b_{J}}$ and $z$-ranges.}
\label{fig:bias_4lum4z}
\end{center}
\end{figure}

Therefore, we can use our $\xi_{Q}(s,20)$ measurements, represented in
Fig. \ref{fig:xi20_4lum4z} and, together with a theoretical estimate of
$\xi_{\rho}(r,20)$, determine the bias that corresponds to that
theoretical assumption and the observed clustering measurements, on the
assumption of a cosmological model. Our results are shown in Fig.
\ref{fig:bias_4lum4z}. To estimate $\xi_{\rho}(r,20)$, we use the $P(k)$
non-linear estimate of \cite{smith}. To determine $\xi_{\rho}(r)$ we
Fourier transform this $P(k)$ estimate, and integrate the result up to
$s \le 20 \Mpch$ to compute $\xi_{\rho}(r,20)$. The parameters used to
generate the $P(k)$ model were: $\Omega_{m}^{0} = 0.3$,
$\Omega_{\Lambda}^{0} = 0.7$, $\Gamma = 0.17$ and, for a better
comparison with Croom et al.'s (2005) results, $\sigma_{8} = 0.84$. This
value is consistent with recent studies \citep[e.g.][]{percival,sig8},
even though recent measurements also tend to suggest somewhat lower
values \citep{wmap2}.

The stars in Fig. \ref{fig:bias_4lum4z} represent the $b$ estimates for
the magnitude-integrated samples, corresponding to the shaded areas in
Fig. \ref{fig:xi20_4lum4z}. These values are very much in agreement with
those found by \cite{scottnew}, using a similar method. The dashed line
is the empirical description of $$b(z) = 0.53+0.289(1+z)^2$$ found by those authors. 

%The dash-dotted line is the $b(z)$ empirical description found by \citeasnoun{peder}, which is given by $b(z)\sigma_{8}= 1+(2(1+z)/5)^5$. In conformity with our previous assumption, we used $\sigma_{8}=0.84$ to determine this $b(z)$ estimate. This description of the bias evolution was determined through measuring the bias from the 2QZ $w_{p}(\sigma)$, and hence using a different statistical analysis. Perhaps for that reason it does not constitute as good a description of the present measurements as that of \citeasnoun{scottnew}. 

The circles refer to our measurements in different magnitude bins. The
red ones correspond to the faintest, $M_{b_{J}}>-23.5$ QSOs; the blue
ones to the $-24.5<M_{b_{J}}<-23.5$ range; the green circles represent
the QSOs with $-25.5<M_{b_{J}}<-24.5$ and the brightest,
$M_{b_{J}}<-25.5$ QSOs are represented by the yellow circles. Given the
size of the error bars, which are related to the errors on the
associated $\xi(20)$ measurements, no categorical conclusion can be
drawn, regarding the possibility of a luminosity-dependent QSO bias. The
uprise in the bias values with redshift is unrelated to the different
QSO luminosities, as a somewhat positive trend occurs for all QSOs
irrespective of their magnitude. This is not entirely true for the
brightest, $M_{b_{J}}<-25.5$, QSOs, (in yellow) for which the bias at
$z\sim 1.3$ seems higher than at higher redshifts. However, given the
small number of QSOs ($355$) within that redshift/magnitude range, this
result would need further study.

The $b$ values for each magnitude are centred in the median redshift of
the QSO sub-sample from which $b$ was determined. Hence, in each
redshift bin, the $z$-displacement of different magnitude points is due
to the non-uniform distribution of the QSOs in the $[M_{b_{J}},z]$
plane. That $z$-displacement, together with the colour-code on the left
side of the plot, makes it easy to relate Fig. \ref{fig:xi20_4lum4z} to
Fig. \ref{fig:bias_4lum4z}.

The red, 2SLAQ-dominated, fainter bin at $z\approx1$ with a relatively
small error bar deviates significantly from the empirical model of
\cite{scottnew}. However, the overall trend is conservatively consistent
with a luminosity-independent QSO bias.\\

The bias of the QSOs is related to the mass of the dark matter halo they
inhabit. In a Gaussian random field the higher the fluctuation threshold
the higher the clustering amplitude of fluctuations. Therefore, by
measuring the clustering of QSOs we can infer the mass of the haloes the
QSOs inhabit. The formalism relating bias and halo mass was firstly
developed by \cite{mowhite}, who assumed a spherical collapse model.
This was then extended to more complicated geometries, such as
ellipsoidal collapse, by \cite{smt01}. In the analysis in this work the
latter will be the adopted formalism. According to these authors, the
bias can be related to the dark halo mass by:

\begin{eqnarray}
b(M_{\mathrm{DMH}},z)& = & 1+\left.\frac{1}{\sqrt{a}\delta_{c}(z)} \right[ \sqrt{a}(a\nu^{2})+\sqrt{a}b(a\nu^{2})^{1-c}\nonumber
            \\
            & - & \left.\frac{(a\nu^{2})^{c}}{(a\nu^{2})^{c}+b(1-c)(1-c/2)}\right]
\label{equation:ellipsoidal_bias}
\end{eqnarray}

with $a=0.707$, $b=0.5$ and $c=0.6$. $\nu$ is defined as $\nu=\delta_{c}(z)/\sigma(M_{\mathrm{DMH}},z)$. $\delta_{c}$ is the critical density for collapse, and is given by: $\delta_{c}= 0.15(12\pi)^{2/3}\Omega_{m}(z)^{0.0055}$ \citep{nfw}. $\sigma(M_{\mathrm{DMH}},z)= \sigma(M_{\mathrm{DMH}})G(z)$, where $\sigma(M_{\mathrm{DMH}})$ is the {\it rms} fluctuation of the density field on the mass scale with value $M_{\mathrm{DMH}}$ and $G(z)$ is the linear growth factor \citep{peebles84,cptlambda}. $\sigma(M_{\mathrm{DMH}})$ can hence be computed as:

\be
\sigma(M_{\mathrm{DMH}})^{2} = \frac{1}{2\pi^{2}}\int_{0}^{\infty}k^{2}P(k)w(kr)^{2}dk
\label{equation:sig_m_int}
\ee
where $P(k)$ is the power spectrum of density perturbations and $w(kr)$ is the Fourier transform of a spherical top hat, which can be given by \citep{peebles}:

\be
w(kr)=3\frac{\sin(kr)-kr\cos(kr)}{(kr)^{3}}
\label{equation:w_peebles}
\ee
where the radius $r$ is related to the mass by:

\be
r=\left(\frac{3 M_{\mathrm{DMH}}}{4\pi \rho_{0}}\right)^{1/3},
\label{equation:radius_mass}
\ee
and $\rho_{0} = \Omega_{m}^{0}\rho_{crit}^{0}$ is the present mean density of the Universe, given by $\rho_{0} = 2.78\times 10^{11} \Omega_{m}^{0} h^{2} M_{\odot}$ Mpc$^{-3}$.

Here, we adopt a linear form of the power spectrum, $P(k) =
P_{0}T(k)^{2}k^{n}$, where $P_{0}$ is simply a normalisation parameter
that depends on $\sigma_{8}$ and $T(k)$ is the transfer function, which
we describe through the analytical formula of \cite{bardeen}.

\begin{figure}
\begin{center}
\centerline{\epsfxsize = 9.0cm
\epsfbox{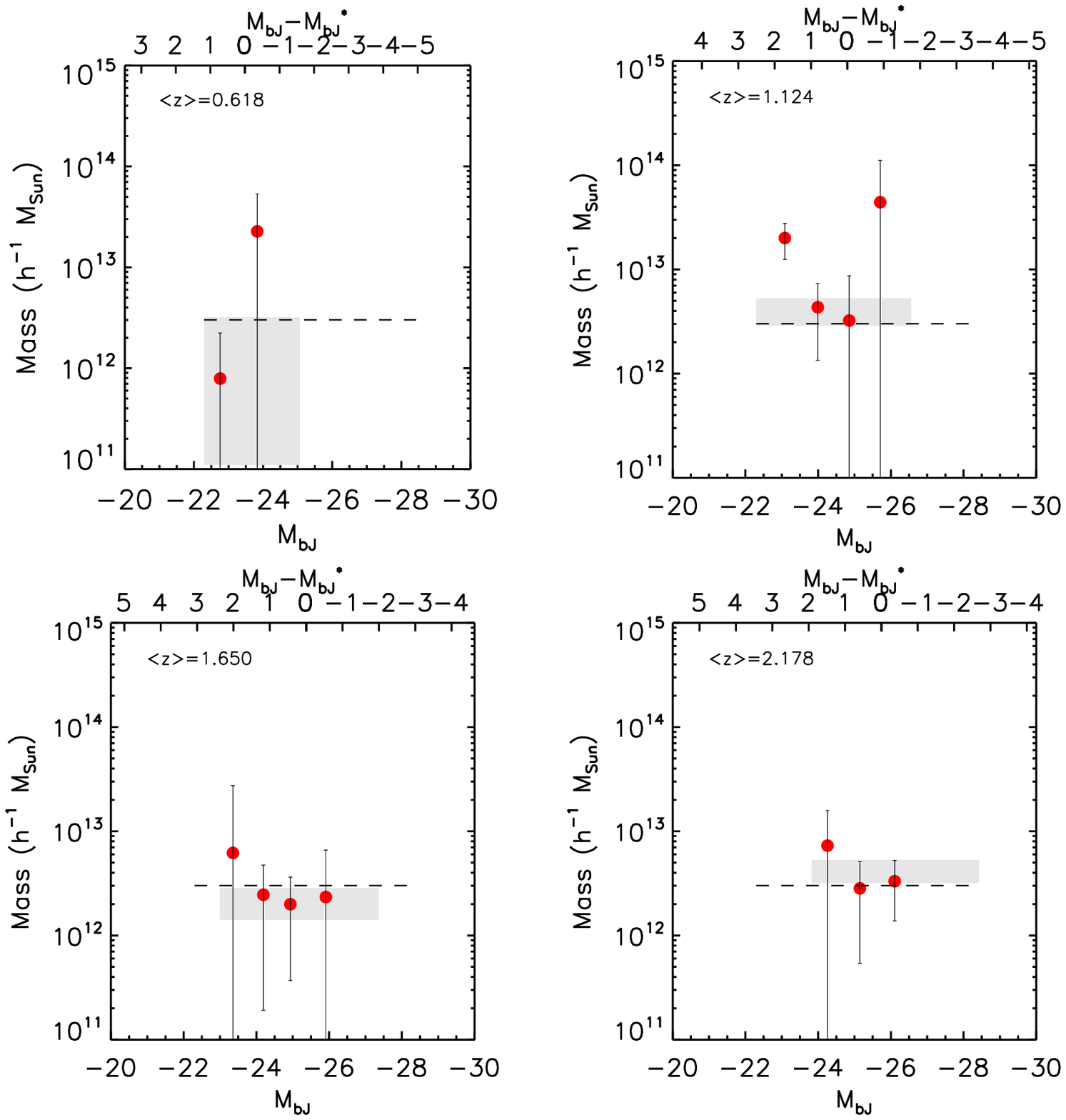}}
\caption{The four panels show the $M_{\mathrm{DMH}}$ estimates in different redshift bins. The median redshift of each $z$-interval is indicated in the top left of each graph. The top horizontal axis shows the magnitude difference relative to $M_{b_{J}}^{*}(<z>)$. The red circles show the dark matter halo mass measurements in different absolute magnitude bins, and are centred on the median values of each bin. The shaded area is the $1 \sigma$ interval for the $M_{\mathrm{DMH}}$ value of all QSOs in that specific redshift interval. The horizontal length of the shaded area represents the range of $M_{b_{J}}$ values for the QSOs in the redshift interval. The dashed line shows the average $M_{\mathrm{DMH}}$ at all redshifts.}
\label{fig:M_DMH_4lum4z}
\end{center}
\end{figure}

The results of performing this analysis using our determination of the
bias is shown in Fig. \ref{fig:M_DMH_4lum4z}. The panels show the dark
matter halo mass associated with different luminosity QSOs, in the same
redshift intervals as those plotted in Fig. \ref{fig:xi20_4lum4z}. The
horizontal axes show the QSO absolute magnitude (bottom), and its
difference relative to $M_{b_{J}}^{*}$ (top axis), similarly to Fig.
\ref{fig:xi20_4lum4z}. In each panel, the red circles represent the
$M_{\mathrm{DMH}}$ measurements in different magnitude bins, with error
bars being the uncertainties corresponding to those obtained in our
previous $b(z)$ estimates. The increase in the relative errors in Fig.
\ref{fig:M_DMH_4lum4z} as compared to Figs. \ref{fig:xi20_4lum4z},
{\ref{fig:bias_4lum4z} is due to the relatively flat slope of the
$\sigma(M_{DMH})$ relation for the $\Lambda CDM$ model as obtained from equn.
\ref{equation:sig_m_int}. The shaded areas represent the $1 \sigma$
$M_{\mathrm{DMH}}$ confidence levels when estimating the masses
associated with all QSOs, irrespective of their luminosities.

We find that, at all redshifts, QSOs seem to inhabit $M_{\mathrm{DMH}}
\sim 3\times 10^{12} h^{-1} \Msun$ haloes (dashed line), very much
in agreement with what was found by \cite{scottnew}. As pointed out by
those authors, this result appears to disfavour the picture of a long-lived QSO
population. As the dark matter halo masses grow, with decreasing
redshift, we would expect to see lower-$z$ QSOs in more massive haloes,
if that were the case. The fact that we do not, means that at
consecutive redshift intervals, we may not be  observing the same QSO
population, but rather distinct sets of objects. However, this
conclusion is based on the 2QZ and 2SLAQ results alone and so the caveat
made at the end of Section 5  about the  higher clustering amplitudes
measured by other authors  for  low redshift AGN still applies.

We also find, through our results, no evidence for $M_{\mathrm{DMH}}$
segregation with QSO magnitude at fixed redshift. All the values seem to
be consistent with a flat $M_{\mathrm{DMH}}-M_{b_{J}}$ trend, indicating
that QSOs seem to live in  $\sim 10^{12} h^{-1} \Msun$ haloes,
independently of their luminosity. This behaviour is inconsistent with
simple, `high peaks', models of QSO biasing where rare, luminous QSOs
might be expected to occupy higher mass haloes.

\section{Estimating black hole masses for different luminosity QSOs}
\label{sec:Lz_massbh}

Several models and theoretical studies have been developed to try to
determine the relation between the mass of the dark matter halo and the
mass of the black holes associated with the observed QSOs. Here we will
consider the two possible evolutionary scenarios considered by
\cite{wyithe}, both based on the results of \cite{ferrarese}: {\bf 1.} a
correlation exists between the dark matter halo mass
($M_{\mathrm{DMH}}$) and the black hole mass ($M_{\mathrm{BH}}$)
\citep{ferrarese} and this relation is unevolving with redshift; {\bf
2.} instead, the correlation between the bulge velocity dispersion (or
circular velocity) and the black hole mass
\citep{ferrarese_merrit,gebhardt}  is assumed to be unevolving with
redshift. We can then estimate the black hole masses associated with
different luminosity QSOs, given that we know the mass of the haloes
that they inhabit, and thus determine if indeed more luminous QSOs are
associated with more massive black holes. For each of these  two
evolutionary scenarios, and following \cite{ferrarese} and
\cite{scottnew}, we will consider three possibilities for the dark
matter halo profile, which affect each of assumed scenarios differently.
We will consider: {\bf a)} an isothermal dark matter profile; {\bf b)} a
NFW \citep{nfw} profile and {\bf c)} a profile inferred from weak
lensing studies \citep{seljak02}, which, for the sake of simplicity, we
will refer to as the ``lensing'' profile.

When assuming a $z$-independent $M_{\mathrm{BH}} - M_{\mathrm{DMH}}$
correlation, the three possible ({\bf a)}, {\bf b)} and {\bf c)}) halo
profiles correspond to the following relations \citep{ferrarese}:\\

{\bf 1. a)} Isothermal profile:
\be
\frac{M_{\mathrm{BH}}}{10^{8}\Msun}\sim 0.027 \left(\frac{M_{\mathrm{DMH}}}{10^{12}\Msun}\right)^{1.82}
\label{equation:iso}
\ee

{\bf 1. b)} NFW profile:
\be
\frac{M_{\mathrm{BH}}}{10^{8}\Msun}\sim 0.1 \left(\frac{M_{\mathrm{DMH}}}{10^{12}\Msun}\right)^{1.65}
\label{equation:nfw}
\ee

{\bf 1. c)} ``Lensing'' profile:
\be
\frac{M_{\mathrm{BH}}}{10^{8}\Msun}\sim 0.67 \left(\frac{M_{\mathrm{DMH}}}{10^{12}\Msun}\right)^{1.82}
\label{equation:lens}
\ee

%These correlations are, analytically, quite similar. The relation between them is actually very close to a simple ``scaling'', with a normalisation parameter ranging from $\sim 10^{-2}$ ({\bf 1. a)}) to $\sim 1$ ({\bf 1. c)}).

%These three possible solutions essentially only differ by a factor which
%``normalises'' the relation between $M_{\mathrm{BH}}$ and
%$M_{\mathrm{DMH}}$, and ranges from $0.027$ (for an isothermal profile)
%to $0.67$ (for a ``lensing'' profile).

If we assume a $z$-independent correlation between the black hole mass
and the circular velocity in the associated bulges (\citealt{shields03},
{\bf 2.}), then other relations are obtained. Following \cite{scottnew}
and \cite{wyithe}, the equivalent relations between the dark matter halo
mass and the black hole mass are given by:

\be
M_{\mathrm{BH}} = e \left(\frac{M_{\mathrm{DMH}}}{10^{12}\Msun}\right)^{2/3}\left(\frac{\Delta \Omega_{m}^{0}}{18\pi^{2}\Omega_{m}(z)}\right)^{5/6}(1+z)^{5/2}
\label{equation:mbh2}
\ee
where $\Delta$ has the form:

\be
\Delta = 18\pi^{2}+82\left(\Omega_{m}(z)-1\right)-39\left(\Omega_{m}(z)-1\right)^{2}
\label{equation:deltabla}
\ee

The constant $e$ is related to the halo density profile. Different
values of $e$ will correspond to the same scenarios as considered in
case {\bf 1.}. Hence, and following \cite{wyithe}, we have that:\\

{\bf 2. a)} For an isothermal profile:
\be
e \sim 10^{-5.1}
\label{equation:iso2}
\ee

{\bf 2. b)} For a NFW profile:
\be
e \sim 3.7 \times 10^{-5.1}
\label{equation:nfw2}
\ee

{\bf 2. c)} For the ``lensing'' profile:
\be
e \sim 25 \times 10^{-5.1}
\label{equation:lens2}
\ee

Again, as in case {\bf 1.}, the three different possibilities considered
for the density profile differ only in terms of a normalisation
parameter, in this case, given  by the constant $e$.

We now use relations {\bf 1.} - {\bf 2.}, {\bf a)}, {\bf b)} and {\bf
c)}, to determine the mass of the black holes that correspond to our
$M_{\mathrm{DMH}}$ measurements, under different assumptions, and
determine if, with the current data, we can relate the black hole
mass to the QSO luminosity.

 \begin{figure}
\begin{center}
\centerline{\epsfxsize = 9.0cm
\epsfbox{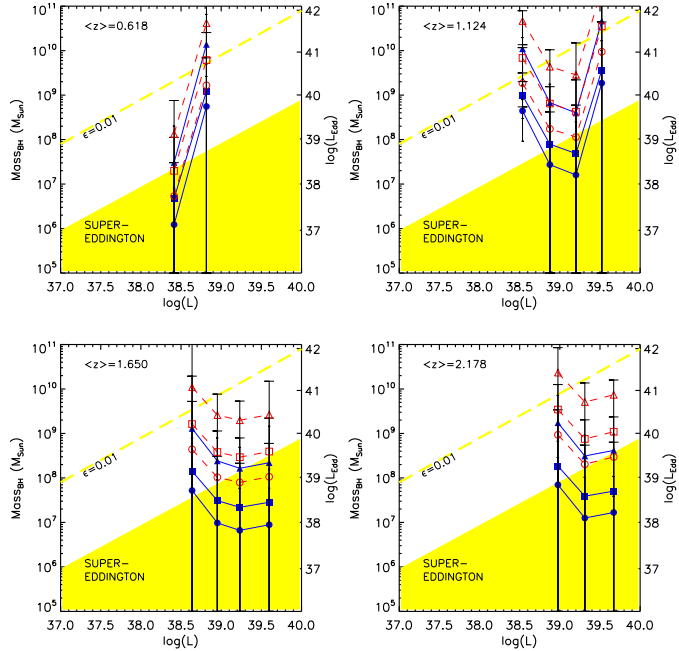}}
\caption{Black hole mass as a function of luminosity, in different
redshift bins. The filled symbols and solid lines are obtained assuming
a $M_{\mathrm{BH}}$ - $M_{\mathrm{DMH}}$ relation which is independent
of $z$. The dashed lines and open symbols, which also correspond to the
errorbars with larger tickmarks, assume a $z$-independent
$M_{\mathrm{BH}}$ - $\sigma_{c}$ relation. In both cases, the circles,
squares and triangles correspond to isothermal, NFW and
lens-studies-based halo density profile, respectively. The points are
located at the median luminosity value of the QSO sub-sample to which they
correspond. On the vertical axis on the right of each panel is the
equivalent Eddington luminosity scale to that for $M_{\mathrm{BH}}$, on
the left. The yellow shaded  area represents the super-Eddington, $L/L_{Edd}>1$,
regime. The dashed yellow line corresponds to a Eddington efficiency
$\epsilon = 0.01$. It can be seen that some models imply super-Eddington
solutions, and hence are unlikely to occur. Most of the models though,
correspond to $0.01 \lsim \epsilon \lsim 1.0$ values.}
\label{fig:M_BH_4lum4z}
\end{center}
\end{figure}

Our results are shown in Fig. \ref{fig:M_BH_4lum4z}. Each panel shows
the results obtained in a given redshift bin. Plotted is the black hole
mass as a function of QSO luminosity. To determine the bolometric
luminosity from $M_{b_{J}}$ we use \citep{scottnew}:

\be
L_{bol} = 10^{\left(79.42-M_{b_{J}}\right)/2.66} W
\label{equation:Lbol}
\ee

The blue filled symbols and solid lines refer to hypothesis {\bf 1.}, where we assume a $M_{\mathrm{BH}}-M_{\mathrm{DMH}}$ $z$-independent relation. The red open symbols and dashed lines relate to hypothesis {\bf 2.}, where we assume a $M_{\mathrm{BH}}-\sigma_{c}$ relation independent of $z$. The filled and open circles show the {\bf a)} estimates, in  Eqs. \ref{equation:iso} and \ref{equation:iso2}, respectively, on which we assume an isothermal density profile. The squares show the results if we assume a NFW profile ({\bf b)}) and the triangles if we assume the lensing profile ({\bf c)}). The error bars are the corresponding uncertainties to those on the $M_{\mathrm{DMH}}$ measurements, plotted in Fig. \ref{fig:M_DMH_4lum4z}. To distinguish between the error bars, the ones that refer to hypothesis {\bf 1.} are represented with short tick marks, whereas the ones that refer to hypothesis {\bf 2.} have longer tick marks.

For both of the assumptions, {\bf 1.} or {\bf 2.}, the dark matter halo ``lensing'' density profile corresponds to more massive black holes, and the isothermal density profile corresponds to the least massive black holes, as expected. Also, it becomes evident that assuming different profiles, being it under $z$-independent $M_{\mathrm{BH}}-M_{\mathrm{DMH}}$ or $M_{\mathrm{BH}}-\sigma_{c}$ scenarios, simply ``shifts'' the $M_{\mathrm{BH}} - \log(L)$ relation vertically. Even though the errors associated with the $M_{\mathrm{BH}}$ are large, we can say that our values are consistent with those of \cite{scottnew}, who studied the evolution of $M_{\mathrm{BH}}$ with redshift.

%The ``curved feature'' seen in the second and third panels is due to a similar trend being observed in our $\xi_{20}$ clustering analysis, where we reported a higher clustering amplitude at the lowest and highest magnitudes than at intermediate luminosities.\\

Also shown, on the right-hand side vertical axis in each panel, is the Eddington luminosity. This is determined directly from the black hole mass as follows:

\be
L_{Edd} = 10^{39.1}\left(\frac{M_{\mathrm{BH}}}{10^{8}\Msun}\right)W
\label{equation:Ledd_mBH}
\ee

The yellow area in the bottom of each panel represents the values of
$M_{\mathrm{BH}}$ that correspond to ``super-Eddington'' solutions ie,
$L/L_{Edd}>1$. The dashed line represents the $M_{\mathrm{BH}} -
\log(L)$ relation for an Eddington efficiency of $\epsilon = L/L_{Edd} =
0.01$. It can be seen that, for some of the scenarios considered, the
mean efficiency is super-Eddington, in particular for models {\bf 1.a)}
and {\bf 1.b)}, ie, assuming an isothermal profile and an NFW profile,
when considering that the $M_{\mathrm{BH}}-M_{\mathrm{DMH}}$ relation
that does not evolve with redshift. One could argue that these relations are therefore
unlikely to occur. Most of the remaining models suggest accretion
efficiencies of $0.01\lsim \epsilon \lsim 1$. It is somewhat unfortunate
that the size of error bars do not allow us to draw conclusions
regarding the significance of potential changes of black hole mass with
luminosity of the associated QSO.\\

We averaged the data over the whole redshift range to test, through a
simple $\chi^{2}$ analysis, the hypothesis that QSOs do not accrete at a
fixed fraction of Eddington. Fig. \ref{fig:M_BH_4lum4ztest} represents
the results, by assuming the ``lensing'' halo density profile and
$z$-independent $M_{\mathrm{BH}}$ - $\sigma_{c}$ relation (open red
triangles). Also shown is the best fitting value of $\epsilon$ for that
assumption.

 \begin{figure}
\begin{center}
\centerline{\epsfxsize = 9.0cm
\epsfbox{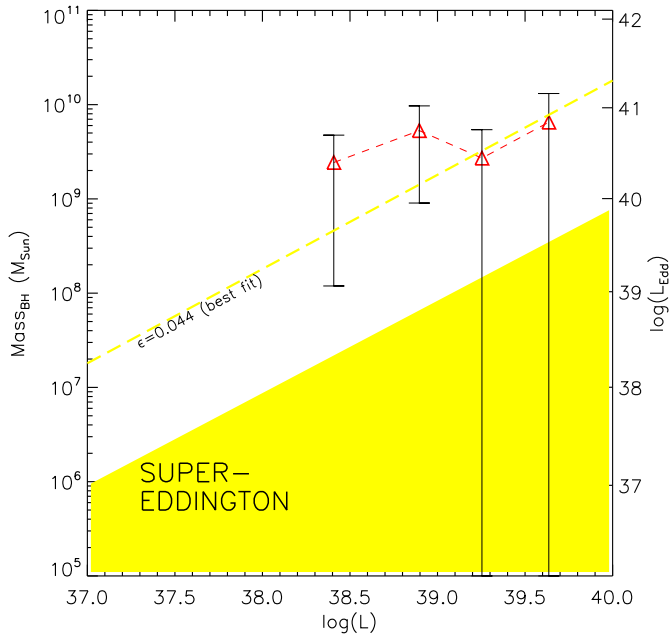}}
\caption{Black hole mass as a function of luminosity, over all redshifts. Here we assume the ``lensing'' halo density profile and that the $M_{\mathrm{BH}}$ - $sigma$ relation is $z$-independent. The best fitting value of $\epsilon$ is shown by the dashed yellow line.}
\label{fig:M_BH_4lum4ztest}
\end{center}
\end{figure}

From the ``flat'' trend  observed in the measured  $M_{\mathrm{BH}}-L$ relation,
black hole mass seems approximately independent of  QSO luminosity. 
However, this does not permit us to exclude the hypothesis that high-$z$
QSOs accrete at a fixed fraction of Eddington, as a model characterised
by a constant value of $\epsilon$ is still a good fit to the data
($\epsilon = 0.044$ with  $\chi^{2} = 1.58$; 3 d.f.). Given recent
studies of \cite{hopkins1} and \cite{lidz06}, who argue that bright and
faint QSOs are similar sources, but observed at different stages of
their activity, one could expect both luminous and faint QSOs to be
associated with equally massive black holes. This would thus lead to
higher values of accretion efficiency for brighter QSOs and lower for
fainter QSOs. Such a model can still be in agreement with the current
analysis, given the ``flat'' trend of the $M_{\mathrm{BH}}$ values as a
function of luminosity.

%Our analysis is supported by a $\chi^{2}$ fit to the measured efficiencies. We  found that, assuming a $z$-independent  $M_{\mathrm{BH}}$ - $M_{\mathrm{DMH}}$ relation, $\epsilon = 0.126_{-0.037}^{+0.056}$, with $\chi^{2}_{reduced} = 4.34$ (3 degrees of freedom), corresponding to a rejection of $99.9954 \%$. If we instead assumed that the $M_{\mathrm{BH}}$ - $\sigma_{c}$ relation does not evolve with redshift, we obtain $\epsilon = 0.015_{-0.004}^{+0.006}$, with $\chi^{2}_{reduced} = 2.10$ (3 degrees of freedom), corresponding to a rejection probability of $99.9017 \%$. Given the $\chi^{2}$ values, it is still possible to reject a model where the QSOs of a given mass accrete at a fixed ratio $\epsilon$. 

Hence, our results show that, if halo mass and black hole mass are
closely correlated, then we cannot reject a model where black hole mass
depends on QSO luminosity and accretion efficiency. It should be noted
that we have assumed that the dispersion in the black hole mass and halo
mass is small. This assumption is supported by the existence of
reasonably tight bulge mass - velocity dispersion relations
\citep{tremaine02}. But clearly, if this assumption proved incorrect,
then the results above would be affected by the high dispersion in the
$M_{bulge}- M_{\mathrm{BH}}$ relation.

\section{Conclusions}
\label{sec:conc}

The 2SLAQ QSO survey is an important tool for QSO clustering studies. 
{\it Firstly}, the 2SLAQ QSO survey complements the previous 2QZ sample
in terms of $z$-space distortion analyses. We have shown that a
double-power law $\xi(r)$ model, which is a good description of the 2QZ
real-space clustering, still describes well both the $z$-space and
projected clustering measurements of the 2QZ and 2SLAQ samples combined.
We fit the dynamical and geometrical distortions of the
$\xi(\sigma,\pi)$ contours, extending the formalism developed by
\cite{hamil92} and \cite{msuto} to include a double-power law $\xi(r)$
model and fitting different ``test'' cosmologies \citep{ap,bph,me2}. We
find that the subsequent confidence levels obtained in $\Omega_{m}^{0}$
and $\beta(z)$ are similar to those obtained when using solely the 2QZ
data, but tighter due to the increased statistics from extra 2SLAQ QSO
pairs, and also the additional cross-correlation pairs in the NGC 2SLAQ
and 2QZ overlapping  volumes. When combining these results with
orthogonal contours obtained from linear theory of density
perturbations, we find that $\Omega_{m}^{0} = 0.25_{-0.07}^{+0.09}$,
$\beta(z) = 0.60_{-0.11}^{+0.14}$, similar to the values obtained from
the 2QZ data alone ($\Omega_{m}^{0} = 0.35_{-0.13}^{+0.19}$, $\beta(z) =
0.50_{-0.15}^{+0.13}$). The new results imply $b(z=1.4)=1.5\pm0.2$ for
the QSO bias.

{\it Secondly}, the 2SLAQ QSO survey constitutes a new dataset with a
potentially central role in terms of breaking the $L$-$z$ degeneracy.
The sample extends $1$ magnitude fainter than the 2QZ, and spans the
same $z$-range. Hence, the combination of both provides a unique
dataset, as the overall magnitude range probed is similar, both at low
and high-$z$. This allows us to interpret clustering results and
possible luminosity-dependent measurements in different redshift bins,
hence reducing any evolutionary biases. Our results are consistent with
luminosity-independent QSO clustering and in agreement with those of
\cite{scottnew}; QSOs seem to inhabit $\sim 3\times 10^{12} h^{-1}\
\Msun$ haloes, independently of their redshift {\em or} luminosity. Our
results do not show a tight correlation between halo mass and QSO
luminosity at fixed redshift, as would be expected from simple "high
peaks" models of QSO biasing  where fainter QSOs populate lower mass
haloes.

%Apart from at the lowest redshift bin where there we do not have very luminous QSOs, all redshift bins indicate a ``curved'' feature in terms of clustering measurements: the faintest and brightest QSOs seem to have a higher clustering amplitude than intermediate luminosity QSOs. This feature however, is not statistically significant. From the QSO bias we estimate the mass of the haloes the 2QZ and 2SLAQ QSOs inhabit. 

Our $M_{\mathrm{DMH}}$ vs. $M_{b_J}$ results  agree with 
the predictions of \cite{lidz06}, whose simulation results based on the models of
\cite{hopkins1,hopkins05, hopkins3, hopkins}
suggest that QSO luminosity may not be correlated with the mass of
the host dark matter halo. The reason is that the same massive haloes
host faint and bright QSOs and the difference in luminosity is due to
the QSOs being observed in different periods of their lifetime. Another
consequence is that QSO clustering should not correlate strongly with
luminosity, again, just as shown by our data.

These authors' analysis also support the results shown in the present
paper and by \cite{scottnew}, and conclude that QSO clustering and halo
mass do not evolve strongly with redshift, even though QSO bias
substantially increases as we move to higher $z$. This could hint at
possible anti-hierarchical QSO formation \citep{merloni,cowie,lidz06},
as haloes harbouring QSOs would have deeper potential wells at high-$z$
than at low-$z$, leading to more luminous black holes being observed at
high-$z$ than at low-$z$. The reason for the rapid decrease of QSO bias
with time is related to haloes of $\sim 10^{12} - 10^{13} \Msun$
corresponding to rarer, high-density-contrast peaks at higher redshift.
The results of those authors also predict that a large range in QSO
luminosity should correspond to a very restricted range in QSO halo
masses, as our observations and measurements seem to indicate.

By assuming different density profiles for the dark matter halo and
$z$-independent relations (such as $M_{\mathrm{BH}}$ -
$M_{\mathrm{DMH}}$ or $M_{\mathrm{BH}}$ - $\sigma_{c}$) we can estimate
the masses of the black holes associated with the QSOs. If the Eddington
limit is a relevant limit for the accretion rate, and if one assumes
that the $M_{\mathrm{BH}}$ - $M_{\mathrm{DMH}}$ relation is
$z$-independent, then isothermal and NFW density profiles are not likely
to be appropriate for the haloes these QSOs inhabit, as they predict
super-Eddington accretions. This is no longer true if one assumes that
the $M_{\mathrm{BH}}$ - $\sigma_{c}$ is independent of redshift,
instead. Most of the other assumptions imply $\sim 10^8$ -- $10^{10}
\Msun$ black holes, and accretion efficiencies of $0.01 \lsim \epsilon
\lsim 1$. Our results suggest that at a given redshift, black hole mass
is not strongly dependent on QSO bolometric luminosity, but a fixed
value for the accretion efficiency is still a good fit to the data.

These results are in agreement with those of \cite{mclure}. In
particular the latter measured the masses and Eddington efficiencies of
high-$z$ black holes using data from the SDSS DR1, through modelling the
QSO spectra. Their analysis, significantly different from that presented
here, results in $M_{\mathrm{BH}}$ and efficiency $\epsilon$ values
similar to those we obtained. Different relations between the black hole
and dark halo masses differ almost by a scaling factor. Therefore, the
trend observed in the $M_{\mathrm{BH}} - \log(L)$ plot is the same
irrespective of the halo density profile and $M_{\mathrm{BH}}$ -
$M_{\mathrm{DMH}}$;$M_{\mathrm{BH}}$ - $\sigma_{c}$ relation.

\section*{Acknowledgments}

We thank Peder Norberg and Carlton Baugh for useful comments and discussions. JA acknowledges financial support from the European Community's Human Potential Program under contract HPRN-CT-2002-00316, SISCO. NPR acknowledges a PPARC Studentship.

We warmly thank all the present and former staff of the
Anglo-Australian Observatory for their work in building and operating
the 2dF facility.  The 2SLAQ Survey is based on
observations made with the Anglo-Australian Telescope and for the SDSS.
Funding for the SDSS and SDSS-II has been
provided by the Alfred P. Sloan Foundation,
the Participating Institutions,
the National Science Foundation,
the U.S. Department of Energy,
the National Aeronautics and Space Administration,
the Japanese Monbukagakusho,
the Max Planck Society,
and the Higher Educatioon Funding Council for England.
The SDSS Web site \hbox{is {\tt http://www.sdss.org/}.}

The SDSS is managed by the Astrophysical Research Consortium
(ARC) for the Participating Institutions.  The Participating
Institutions are
the American Museum of Natural History,
Astrophysical Institute of Potsdam,
University of Basel,
Cambridge University
Case Western Reserve University,
University of Chicago,
Drexel University,
Fermilab,
the Institute for Advanced Study,
the Japan Participation Group,
Johns Hopkins University,
the Joint Insitutue for Nuclear Astrophysics,
the Kavli Insitute for Particle Astrophysics and Cosmology,
the Korean Scientist Group,
the Chinese Academy of Sciences (LAMOST),
Los Alamos National Laboratory,
the Max-Planck-Institute for Astronomy (MPIA),
the Max-Planck-Institute for Astrophysics (MPA),
New Mexico State University,
Ohio State University,
University of Pittsburgh,
University of Portsmouth,
Princeton University,
the United States Naval Observatory,
and the University of Washington.

\bibliographystyle{mn2e}
\bibliography{2slaq_qso_paper_ts1}

\label{lastpage}

\end{document}